\documentclass[aps,prl,amsmath,twocolumn,superscriptaddress,letterpaper,floatfix]{revtex4}
\usepackage{graphicx,color}
\usepackage{verbatim}
\usepackage{amssymb}   % for math
\usepackage{amsmath}
\usepackage{amsfonts}
\usepackage{mathdots}
\usepackage{hyperref}
\usepackage{epsfig}
\usepackage{braket}
\usepackage{bm}
\begin{document}
	\title{Spin-orbit-parity coupled superconductivity in topological monolayer WTe$_2$}
	\author{Ying-Ming Xie} \thanks{These authors contributed equally to this work.}
	\author{Benjamin T. Zhou} \thanks{These authors contributed equally to this work.}
	\author{K. T. Law} \thanks{Corresponding author.\\phlaw@ust.hk}
	\affiliation{Department of Physics, Hong Kong University of Science and Technology, Clear Water Bay, Hong Kong, China} 	
	
	\date{\today}
	\begin{abstract}
		Recent experiments reported gate-induced superconductivity in the monolayer 1T$'$-WTe$_2$ which is a two-dimensional topological insulator in its normal state \cite{Fatemi, Sajadi}. The in-plane upper critical field $B_{c2}$ is found to exceed the conventional Pauli paramagnetic limit $B_p$ by 1-3 times. The enhancement cannot be explained by conventional spin-orbit coupling which vanishes due to inversion symmetry. In this work, we unveil some distinctive superconducting properties of centrosymmetric 1T$'$-WTe$_2$ which arise from the coupling of spin, momentum and band parity degrees of freedom. As a result of this spin-orbit-parity coupling (SOPC): (i) there is a first-order superconductor-metal transition at $B_{c2}$ much higher than the Pauli paramagnetic limit $B_p$, (ii)  spin-susceptibility is anisotropic with respect to in-plane directions and can result in possible anisotropic $B_{c2}$ and (iii) the $B_{c2}$ exhibits a strong gate dependence as the spin-orbit-parity coupling is significant only near the topological band crossing points. The importance of SOPC on the topologically nontrivial inter-orbital pairing phase is also discussed. Our theory generally applies to centrosymmetric materials with topological band inversions.
	\end{abstract}
	
	
\maketitle
{\emph {Introduction.}}--- Recently, centrosymmetric monolayer 1T$'$-structure WTe$_2$, which is a two-dimensional topological insulator with helical edge states \cite{Qian,roberto2016,Tang2017,Fei2017,WuSanfeng}, has been found to become superconducting upon electro-gating \cite{Fatemi, Sajadi}. The coexistence of helical edge states and superconductivity establishes the system as a promising platform to create Majorana fermions \cite{Fu2009, Beenakker} and thus attracts wide on-going attention. Interestingly, the in-plane $B_{c2}$ of the superconducting topological insulator was found to be 1-3 times higher than the usual Pauli paramagnetic limit $B_p$ \cite{Fatemi, Sajadi}. 

It has been well established that spin-orbit couplings which lift spin degeneracies in electronic bands can enhance the $B_{c2}$ in noncentrosymmetric superconductors \cite{Frigeri, Gorkov}. In particular, Ising superconductors such as noncentrosymmetric 2H-structure MoS$_2$, NbSe$_2$ and WS$_2$, have been shown to exhibit in-plane $B_{c2}$ several times higher than $B_p$ due to Ising spin-orbit coupling  \cite{Lu1353, Xi2015, Saito2016, delaBarrera2018, Ye, JianWang, Sohn2018, Law, Tewari, Aji, He2018, Meyer}. Despite similar chemical compositions and layered structures, 1T$'$-structure WTe$_2$ respects inversion symmetry such that spin-orbit coupling terms which involve only spin and momentum degrees of freedom are forbidden \cite{Qian,Tang2017,roberto2016}. Therefore, the mechanism behind the observed enhancement of $B_{c2}$ remains unknown. 

\begin{figure}
	\centering
	\includegraphics[width=1\linewidth]{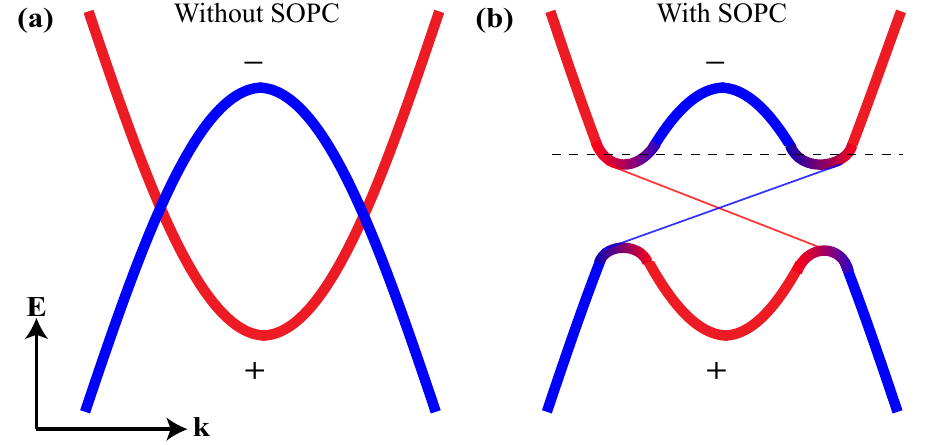}
	\caption{Schematic band structure of two inverted bands without spin-orbit-parity coupling (SOPC) (a) and with SOPC coupling (b). The $+ (-)$ sign labels the even (odd) parity of the band. Bands with even and odd parities in 1T$'$-WTe$_2$ originate predominantly from the $d-$ and $p-$ atomic orbitals respectively. In (b), the SOPC opens a topologically nontrivial gap at the band crossing points and edge states emerge (thin lines in the gap). Only states close to the crossing points with heavily mixed orbital parities can experience strong SOPC. The horizontal dashed line in (b) denotes the chemical potential at which superconductivity is observed in the experiment. }
	\label{fig1}
\end{figure}

In this work, we show that inversion symmetry allows the spin, momentum and parities of the electronic states to couple in 1T$'$-WTe$_2$. We refer to this coupling as spin-orbit-parity coupling (SOPC). The SOPC not only opens a topological gap (as depicted in Fig\ref{fig1}), and creates the helical edge modes \cite{Qian,roberto2016,Tang2017,Fei2017,WuSanfeng}, but also pins the electron spins and renormalizes the effect of external Zeeman fields to enhance the $B_{c2}$. Importantly, the SOPC dramatically affects the superconducting properties such that: (i) 1T$'$-WTe$_2$  undergoes a first-order superconductor-metal transition at $B_{c2}$, similar to conventional $s$-wave superconductors \cite{maki}. However, the transition happens at a much higher field than $B_p$; (ii) the spin susceptibility and $B_{c2}$ can be anisotropic with respect to in-plane magnetic field directions; (iii) the $B_{c2}$ is strongly gate-dependent as the SOPC is effective only for states near the topological band crossing points (band crossing involving bands with opposite parities). These properties distinguish superconductors with SOPC from noncentrosymmetric and convensional $s$-wave superconductors. Comparison among superconductors with SOPC, Ising superconductors and conventional $s$-wave superconductors is presented in Table \ref{table:01}.

Importantly, SOPC widely exists in topological materials such as superconducting Cu-dopped Bi$_2$Se$_3$ \cite{Zhang2009,Chaoxing,Fuliang,Hashimoto}. However, orbital depairing effects in three-dimensional materials overwhelm the Zeeman effect in the superconducting state. Moreover, superconductivity in Cu-doped Bi$_2$Se$_3$ sets in when the chemical potential lies high above the band crossing points where the SOPC effect is weak \cite{ong,Hasan}. Therefore, atomically thin 1T$'$-WTe$_2$, being superconducting near the band crossing points as depicted in Fig.\ref{fig1}b, provides an ideal platform to study spin-orbit-parity coupled superconductivity. Interestingly, we further show that SOPC is important for stablizing the inter-orbital pairing phases which can be topologically non-trivial.

Moreover, an enhanced $B_{c2}$ has been observed in centrosymmetric monolayer 1T$'$-MoTe$_2$ \cite{Noah2019}, which was attributed to Rashba spin-orbit coupling due to gate-induced inversion breaking. Our theory suggests that the $B_{c2}$ enhancement in 1T$'$-MoTe$_2$ can be readily explained by the SOPC and inversion breaking is inessential. 

\begin{figure}
	\centering
	\includegraphics[width=1\linewidth]{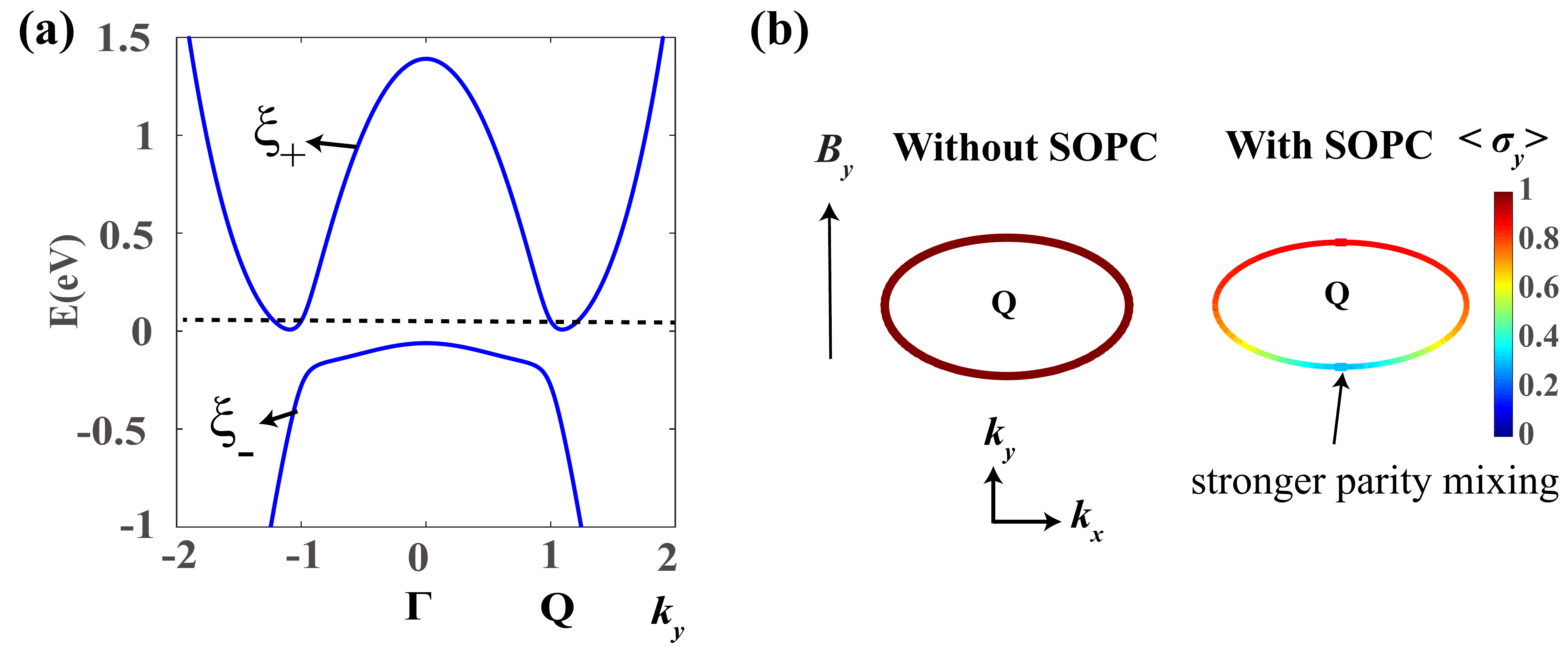}
	\caption{(a) Normal-state band structure of monolayer WTe$_2$. Hybridyzation between $p$- and $d$-bands from SOPC opens a topologically nontrivial gap near $\pm Q$ and results in two $Q$-valleys in the conduction bands. (b) Expectation value of spin-$y$ component $\braket{\sigma_y}$ without(left)/with(right) SOPC on the Fermi surface contours under a weak Zeeman field $\bm{B}=B_y\hat{y}$ (Zeeman strength $\sim 1$ meV, contours around $+Q$ is shown here). The net spin along $y$-direction induced by $B_y$ is reduced by the pinning due to SOPC.}
	\label{fig2}
\end{figure}

{\emph{Model Hamiltonian of superconducting monolayer 1T$'$-WTe$_2$.}}---The symmetry group of a monolayer 1T$'$-WTe$_2$ is generated by time-reversal, one in-plane mirror symmetry, and spatial inversion. These symmetries dictate the form of a four-band $\bm{k}\cdot\bm{p}$ Hamiltonian which describes the normal state of WTe$_2$ \cite{Qian, NoteX}:
\begin{align}\label{eq:normalH}
H_0(\bm{k})=&\epsilon_0(\bm{k})+\mathcal{M}(\bm{k})s_z+vk_xs_y\nonumber+A_xk_xs_x\sigma_y\\
&+A_yk_ys_x\sigma_x+A_zk_ys_x\sigma_z,
\end{align} 
where $\epsilon_0(\bm{k})=t_x^{+}k_x^2+t_y^{+}k_y^2+\frac{1}{2}t'_{x}k_y^4+\frac{1}{2}t'_{y}k_y^4-\mu$,  $\mathcal{M}(\bm{k})=-\delta+t_x^{-}k_x^2+t_y^{-}k_y^2-\frac{1}{2}t'_xk_x^4-\frac{1}{2}t'_{y}k_y^4$. Here, the $s$-matrices operate on the orbital degrees of freedom formed by ($p,d$)-orbitals with opposite parities, and $\sigma$-matrices act on the spin space. Notably, $\delta$ determines the order of the band at $\bm{k}=0$. When $\delta>0$, there is a band inversion while the SOPC terms open a topologically non-trivial gap and the system become a topological insulator as schematically depicted in Fig.\ref{fig1}b. Derivation of the symmetry allowed terms and the model parameters are given in the Supplementary Materials \cite{NoteX}. In $H_0$, the energy dispersions of the bands are given by $\xi_{\pm}(\bm{k})$ (as shown in Fig.\ref{fig2}a), with each band being two-fold degenerate due to both the spatial inversion and time-reversal symmetries.

We emphasize that the usual spin-orbit coupling terms which involve $\bm{k}$ and $\sigma$ only are forbidden by inversion symmetry. However, it is possible to have an SOPC term $\bm{\hat{g}\cdot}\bm{\sigma}$, where $\bm{\hat{g}}=(A_yk_y,A_xk_x,A_zk_y)s_x$. Importantly, the SOPC term is proportional to $s_x$ and $\braket{ \Psi (\bm{k})| \bm{\hat{g}\cdot}\bm{\sigma} |\Psi (\bm{k})}$ is significant only for $\Psi$ with strongly hybridized $p-$ and $d-$orbitals. This happens only near the topological band crossing points as schematically depicted in Fig.\ref{fig1}b. Interestingly, superconductivity in 1T$'$-WTe$_2$ was observed experimentally when conduction band states near the band crossing points at $\pm Q$ are filled (Fig.\ref{fig2}a) with charge density $n\sim 10^{13} cm^{-2}$ \cite{Fatemi, Sajadi}. Thus, 1T$'$-WTe$_2$ is an ideal platform to study the effects of SOPC on superconductivity. 

Assuming on-site attractive interactions to be dominant, the intra-orbital singlet-pairing phase is expected to be energetically favorable. In this case, the superconducting state under an in-plane magnetic field $\bm{B}$ can be described by the Bogoliubov–de Gennes Hamiltonian:
\begin{equation}\label{eq:HBdG}
H_{BdG}(\bm{k})=H_0(\bm{k})\eta_3+\frac{1}{2}g_su_B\bm{B\cdot\sigma}+\Delta\eta_1,
\end{equation} 
where $\eta$ operates on particle-hole space, $u_B$ is the Bohr magneton, $g_s=2$ is the Landé g factor. 

To understand how SOPC affects the magnetic response to an external Zeeman field, it is instructive to project $H_{BdG}(\bm{k})$ to a manifestly covariant pseudospin basis (MCPB) $\{\ket{\bm{k},\alpha},\ket{\bm{k},\beta}\}$ \cite{Yip,Liang,Venderbos} for the conduction band with energy $\xi_{+}(\bm{k})$, where superconducting pairing is formed. The transformation properties of the MCPB basis can be found in the Supplementary Materials \cite{NoteX}. By projecting $H_{BdG}(\bm{k})$ into the subspace $(\psi^{\dagger}_{\bm{k},\alpha},\psi^{\dagger}_{\bm{k},\beta},\psi_{-\bm{k},\beta},-\psi_{-\bm{k},\alpha})$, the effective pairing Hamiltonian has the form:
\begin{equation}
H_{\text{eff}}(\bm{k})=\xi_{\bm{k}}\eta_3+\frac{1}{2}g_su_B\bm{B\cdot\tilde{\sigma}}(\bm{k})+\Delta \eta_1,
\label{eq:effHam}
\end{equation}
where $\tilde{\sigma}_{i}^{l,l'}(\bm{k})=\braket{\bm{k},l|\sigma_i|\bm{k},l'}=\sum_{j}a_{ij}(\bm{k})\rho_j^{l,l'}$ ($\rho_j$: Pauli matrix in the pseudospin basis) is the projected spin operator in the pseudospin subspace, and the effect of SOPC on electron spins are encoded in the coefficients $a_{ij}(\bm{k})$ (see Supplementary Material \cite{NoteX} for explicit forms of $a_{ij}(\bm{k})$). It is clear from Eq.\ref{eq:effHam} that the Zeeman effect due to external magnetic fields is renormalized by the SOPC term. %To be specific, consider the $A_{y} k_y$ term only and with an applied magnetic field along the $y$-direction, we have $B_y\tilde{\sigma_{y}}=B_y \sqrt{1-A_{y}^2 k_{y}^2/(\mathcal{M}^2(\bm{k})+v^2k_x^2+A_y^2k_y^2)} \rho_{y}$. Therefore, the Zeeman effect is reduced by the SOPC. %This renormalization effect reduces the normal state spin susceptibility and the enhances the $B_{c2}$ in the superconducting state as discussed in details in the next section.

 \begin{figure}
 	\centering
 	\includegraphics[width=1\linewidth]{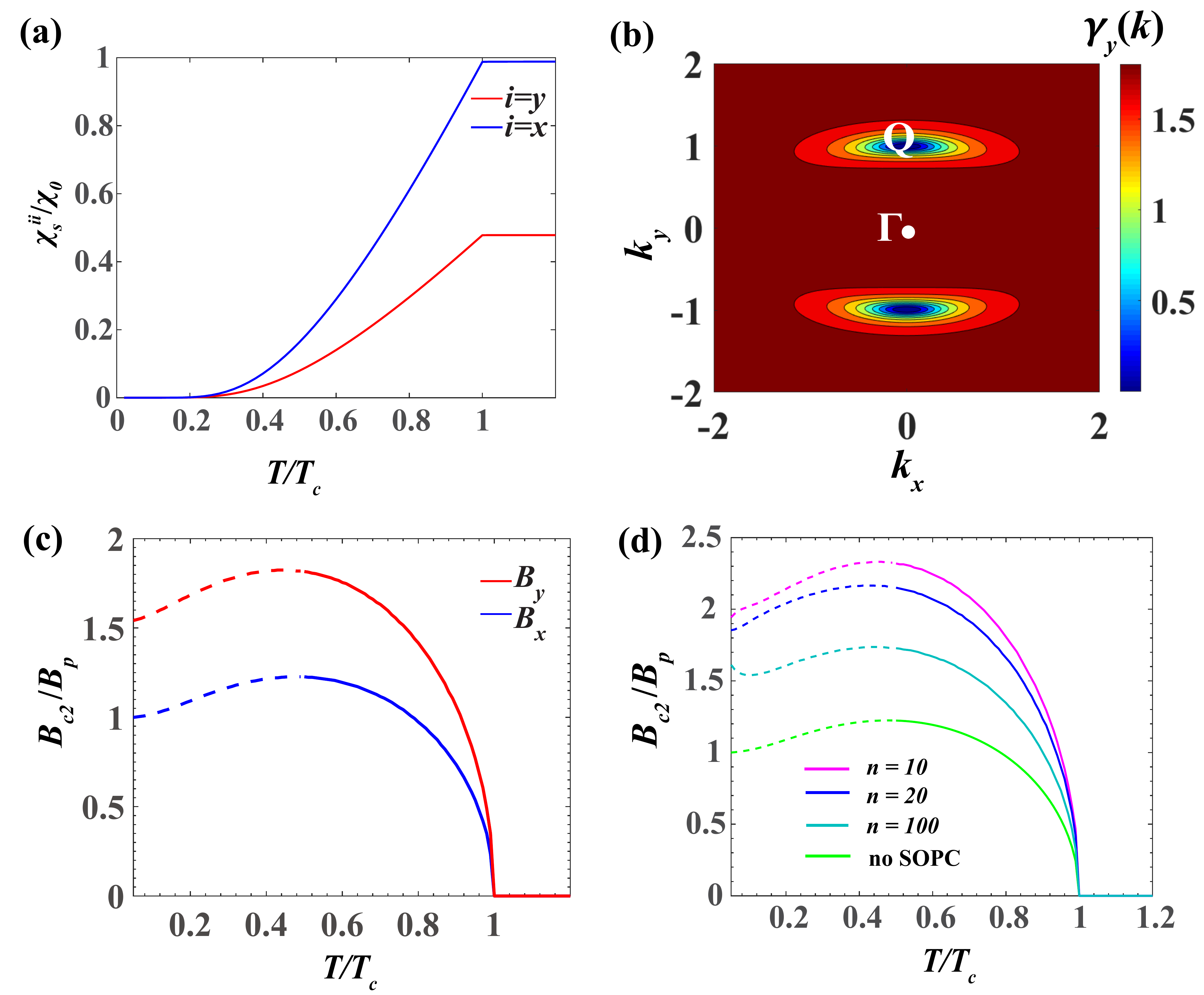}
 	\caption{Enhancement of $B_{c2}$ via SOPC for 1T$'$-WTe$_2$.   (a) Spin susceptibility $\chi_n^{ii}$ ($i=x,y$) as a function of temperature $T$, where the SOPC strength is $A_y=0.855$ eV$\cdot$\AA, Fermi energy $E_F=100$ meV. We set $T_{c}=1$ K according to experimental observations. (b) Value of $\gamma_y$ at different $\bm{k}$. $\gamma_y(\bm{k})$ approaches zero near the band minimum at $\pm Q$.  (c) $B_{c2} - T_{c}$ curves for $\bm{B} = B_x \hat{x}$(blue) and $\bm{B} = B_y \hat{y}$(red). Other parameters are the same as in (a).  (d) $B_{c2} - T_{c}$ curves  for $\bm{B} = B_y \hat{y}$ with different carrier density $n$ in units of $10^{12}$ cm$^{-2}$ and $A_y=1.71$ eV$\cdot$\AA. The case without SOPC (light green curve) is presented for reference. }
 	\label{fig3}
 \end{figure}

To demonstrate the renormalization and the spin-pinning effect encoded in $a_{ij}(\bm{k})$, we assume a weak Zeeman field $\bm{B} = B_y \hat{y}$ in $H_{\text{eff}}$ and plot the Zeeman field induced spin expectation value in the $y$-direction $\braket{\sigma_y}$ for states near the $Q$-point with and without SOPC in Fig.\ref{fig2}b. Evidently, without SOPC, spins along the Fermi surface contours can freely align with $B_y$. In contrast, in the presence of SOPC, spins at different $\bm{k}$ are pinned predominantly to the $x$-direction as the $A_yk_ys_x\sigma_x$ term dominates \cite{NoteX}. It is important to note that in Fig.\ref{fig2}b, the spin pinning is much stronger for states with smaller $k_y$ near the band crossing point due to the stronger mixing between $p-$ and $d-$orbitals in these states. This clearly demonstrates the SOPC effect is not determined by the spin-orbit coupling part $A_yk_y\sigma_x$ alone, but also largely governed by the parity mixing operator $s_x$. In the next section, we show the important effects of SOPC on $B_{c2}$.

{\emph{Enhancement, anisotropy and gate dependence of in-plane $B_{c2}$. }}---Phenomenologically, the normal-state and superconducting free energy densities due to an external in-plane field $\bm{B}$ ($B=|\bm{B}|$) and pairing can be written as $f_n(B) = -\frac{1}{2}\chi_nB^2$, and $f_s(B) = f_{cond} + f_{spin}$ respectively. Here, $\chi_n/\chi_s$ is the normal-state/superconducting spin susceptibility, $f_{cond} =- \frac{1}{2}N(E_F)\Delta_{0}^2$, with $ \Delta_0 = \Delta(B =0)$, is the zero-field condensation energy with $N(E_F)$ being the density of states at Fermi energy, and $f_{spin} = -\frac{1}{2}\chi_sB^2$ is the spin magnetic energy in the superconducting state. $B_{c2}$ can be estimated by identifying the point $f_n(B) = f_s (B)$, yielding $ B_{c2} \approx B_{p}\sqrt{\chi_0/(\chi_n-\chi_s)}$, where $B_{p} = \Delta_0/(\sqrt{2}\mu_B)$, and $\chi_0=2N(E_F)u_B^2$ is the Pauli spin susceptibility of free electron gas. Clearly, $B_{c2}$ can be enhanced to be higher than $B_p$ via: (i) a reduced $\chi_n < \chi_0$, and (ii) a residue $\chi_s \neq 0$.  As shown in the MCPB basis, $H_{\text{eff}}$ has the form of a spin-singlet superconductor, we expect that the superconducting ground state cannot respond to a weak external Zeeman fields, which implies $\chi_{s} = 0$ in the $T \rightarrow 0$ limit.

To demonstrate the vanishing $\chi_{s}$ in WTe$_2$, we calculate the superconducting spin susceptibility $\chi^{ii}_{s}$ ($i=x,y$) given by \cite{Sigrist,abrikosov}:
\begin{eqnarray}
\chi^{ii}_{s}&=&-\frac{1}{2}u_B^2k_BT\sum_{\bm{k}, n}\text{Tr}[\tilde{\sigma_i}\mathcal{G}(\bm{k},i\omega_n)\tilde{\sigma}_i\mathcal{G}(\bm{k},i\omega_n)] \\\nonumber
&=& \frac{1}{2}u_{B}^2\beta\sum_{\bm{k}}\gamma_{i}(\bm{k})\frac{1}{1+\cosh(\beta\sqrt{\xi^2_{\bm{k}}+\Delta^2})},\label{eq:super_sus}
\end{eqnarray}
where $\mathcal{G}(\bm{k},i\omega_n)=(i\omega_n-\xi_{\bm{k}}\eta_3-\Delta\eta_1)^{-1}$ is the Gor'kov Green's function obtained from $H_{\text{eff}}(\bm{k})$ in Eq.\ref{eq:effHam} under zero magnetic field. $T$ is the temperature, $\beta=1/k_BT$, $\omega_n = (2n+1)\pi/k_BT$ denotes the fermionic Matsubara frequency. $\gamma_{i}(\bm{k})=2\sum_{j}a^2_{ij}(\bm{k})$ characterizes the renormalization effect on spins due to SOPC. Clearly, the denominator in the summand in Eq.~(\ref{eq:super_sus}) diverges as $T \rightarrow 0$ due to a finite superconducting gap $\Delta$, thus $\chi^{ii}_{s} (T\rightarrow0) = 0 $ (Fig.\ref{fig3}a). 

%An equivalent form of $\chi_s^{ii}$ is written as $\chi_s^{ii}/\chi_n^{ii}=1-\pi k_BT\sum_n\Delta^2/(\Delta^2+\omega_n^2)^{3/2}$, which is obtained by doing the momentum integral first instead. Later, we will show how this form is modified by the impurity scattering, which turns out that only an additional factor is dressed. %

The vanishing $\chi^{ii}_{s}$ leaves us with the mechanism of enhanced $B_{c2}$ via reduced $\chi_n$. Note that $\chi_n$ is directly given by $\chi_s(\Delta=0)$ in Eq.~(\ref{eq:super_sus}), \textit{i.e.},
\begin{equation}
\chi_{n}^{ii}=\frac{1}{2}u_{B}^2\beta\sum_{\bm{k}}\frac{\gamma_{i}(\bm{k})}{1+\cosh(\beta\xi_{\bm{k}})}=u_B^2N(E_F)\gamma_i (E_F), \label{eq:sus_normal}
\end{equation}
where $\gamma_i (E_F)=\int d^2\bm{k}\gamma_i(\bm{k})\delta(\xi_{\bm{k}}-E_F)/\int d^2\bm{k}\delta(\xi_{\bm{k}}-E_F)$ is the averaged renormalization factor due to SOPC over the Fermi surface (see Supplementary Material \cite{NoteX}). 

As shown in Eq.\ref{eq:sus_normal}, the normal-state spin susceptibility is given by $\chi_n^{ii}= \gamma_i (E_F) \chi_0/2$, with a renormalization factor $\gamma_i (E_F)/2$ due to SOPC. In the low temperature limit, the in-plane critical field along $i$-direction ($i=x,y$) is directly related to the Pauli limit by $B_{c2}^{ii} = B_{p} \sqrt{\chi_0/\chi_n^{ii}} = B_{p} \sqrt{2 /\gamma_i (E_F) }$, which implies $B_{c2}>B_{p}$ when $\gamma_i (E_F)<2$. 

To show the reduced $\chi_n^{yy}$, we plot $\gamma_{y}(\bm{k})$ in the conduction band (Fig.\ref{fig3}b). Evidently, $\gamma_{y}(\bm{k})<2$ holds throughout the whole Brillouin zone. As a result, $\gamma_y (E_F) < 2$ in general, leading to $\chi_n^{yy} < \chi_0$ as consistent with the result in Fig.\ref{fig3}a(red curve) where $\chi_s^{yy} = \chi_n^{yy} < \chi_0$ for $T>T_c$. 

In contrast, we noticed that $\chi_s^{xx} = \chi_n^{xx} \approx \chi_0$ for $T>T_c$ (blue curve in Fig.\ref{fig3}a). This is because $\bm{B} = B_x \hat{x}$ is collinear with the dominant SOPC term $A_yk_ys_x\sigma_x$ and thus can freely align spins to the $x$-direction. As a result, $B^{yy}_{c2}> B_p$ while $B^{xx}_{c2} \approx B_p$ as shown in the $B_{c2} -T_c$ curves in Fig.\ref{fig3}c obtained by solving the linearized gap equation:
\begin{equation}
\frac{2}{U/V}=k_BT\sum_{\bm{k},n}\text{Tr}[G^{(0)}(\bm{k},i\omega_n)\rho_yG^{(0)T}(-\bm{k},-i\omega_n)\rho_y]. 
\end{equation}
Here, $U$ is electron-phonon interaction strength, $V$ is the sample volume, $G^{(0)}(\bm{k},i\omega_n)$ is the normal state Green's function of $H_{\text{eff}}(\bm{k})$ given in Eq.~\ref{eq:effHam}
 (see Supplementary Materials \cite{NoteX} for details). Thus, our results suggest the $B_{c2}$ of an SOPC superconductor can exhibit a strong anisotropy due to the anisotropic SOPC. This provides a distinctive signature of the possible SOPC origin behind the enhanced $B_{c2}$ which is different from the isotropic $B_{c2}$ and $\chi_s$ in both Ising superconductors and conventional superconductors as summarized in Table \ref{table:01}.

Interestingly, $\gamma_y(\bm{k})$ has a strong $\bm{k}$-dependence (Fig.\ref{fig3}b) with the renormalization being strongest (signified by a strongly reduced value of $\gamma_y(\bm{k})$) near the band crossing points at $\pm Q$. As $E_F$ increases upon gating, outer Fermi circles enclosing $\pm Q$ are accessed and $\gamma_y(\bm{k})$ approaches $\gamma_0 = 2$ for free electron gas. This again reflects the \textit{parity-mixing} nature of SOPC: the spin pinning effect due to SOPC terms is strongest near the band crossing points at $\pm Q$ where the $p$- and $d$-orbitals are strongly mixed. As $\bm{k}$ deviates from $\pm Q$, the pairty mixing becomes weaker and the spin pinning effect is suppressed. Such strong dependence of $\gamma_y(\bm{k})$ on Fermi level implies a strong gate-dependence in $B^{yy}_{c2}$. This is explicitly demonstrated by solving the linearized gap equation at different values of carrier density $n$ (Fig.\ref{fig3}d). Consistently, as $n$ increases, the enhancement of $B_{c2}$ is reduced. 

Notably, for superconductors with SOPC, the low temperature sectors of the $B_{c2}-T_c$ curves obtained by linearized gap equations (dashed segments in the range $0 < T < T_1 \approx 0.5T_c$) do not represent the true values of $B_{c2}$ but the supercooling critical field instead \cite{maki}. As we discuss next, the superconductor-metal transition at $B_{c2}$ in this regime is in fact first-order in nature. 
 
\begin{figure}
	\centering
	\includegraphics[width=1\linewidth]{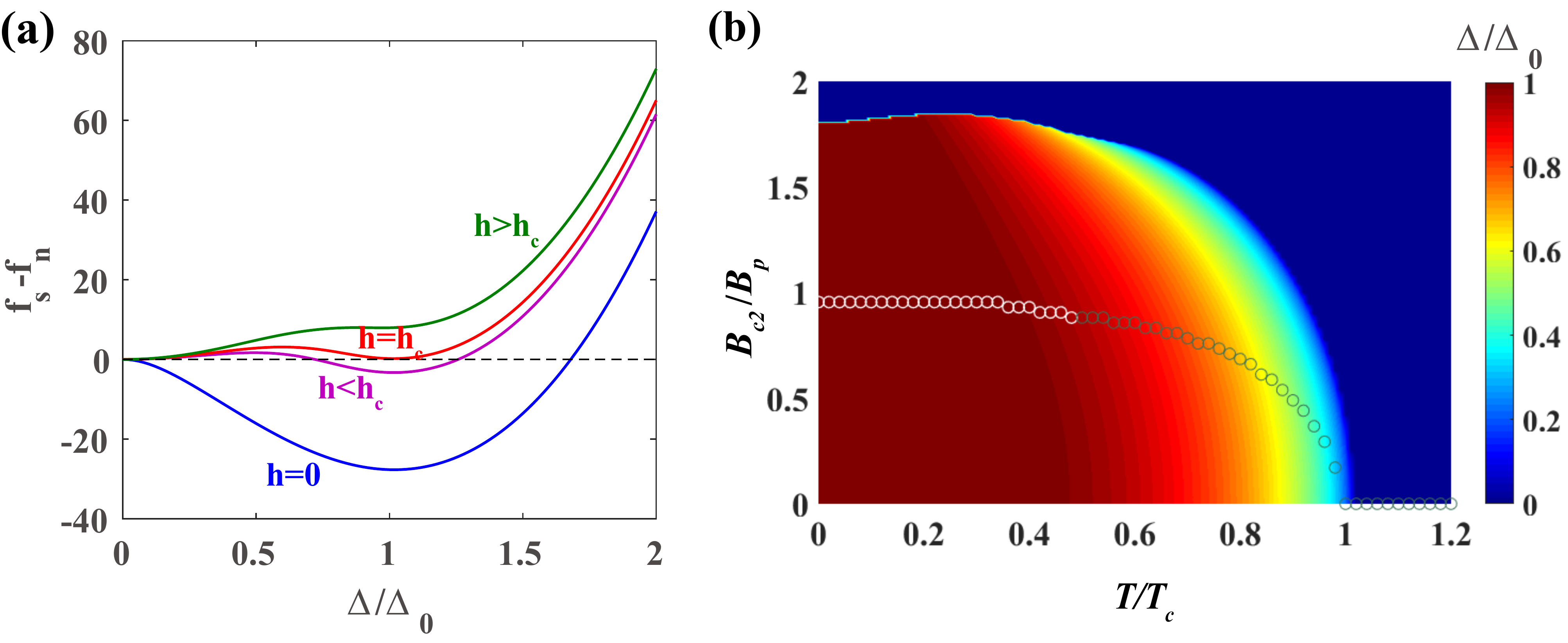}
	\caption{(a) Landscapes of $f_s-f_n$ at $T=0.1 T_c$ in units of meV under $B=0, 1.8B_p, 1.93B_p, 2.2B_p$. $B_{c2}\sim1.93B_p$, with $\Delta_0\approx1.764 k_BT_c$ at $B, T=0$. (b) $B-T$ phase diagram from minimizing $f_s-f_n$ with $A_y=1.71$ eV$\cdot$\AA\ and $n=10\times10^{12}$ cm$^{-2}$. The color represents the magnitude of $\Delta$ at different $B$ and $T$. The line of circles represent the values of $B_{c2}$ in a conventional superconductor, where $B_{c2}(T=0) = B_p$. A first-order transition also occurs in the low temperature regime \cite{maki} (indicated by white circles).}
	\label{fig4}
\end{figure}

\begin{table*}[ht]
\centering
\caption{Comparison among centrosymmetric spin-orbit-parity-coupled(SOPC), Ising and conventional superconductivity.}
\begin{tabular}{c c c c}
\hline \hline
Type of superconductors & \hspace{0.4 in} SOPC  & \hspace{0.4 in} Ising & \hspace{0.4 in} Conventional \tabularnewline
\hline 
Pairing correlations  & \hspace{0.4 in} Singlet & \hspace{0.4 in} Singlet-triplet mixing & \hspace{0.4 in} Singlet  \tabularnewline
\hline
$\chi_s(T=0)$ & \hspace{0.4 in} Zero & \hspace{0.4 in} Finite & \hspace{0.4 in} Zero \tabularnewline
\hline
In-plane $B_{c2}$ & \hspace{0.4 in} $>B_p$ & \hspace{0.4 in}  $>B_p$ & \hspace{0.4 in}  $=B_p$ \tabularnewline
\hline
$B$-driven superconductor-metal transition as $T \rightarrow 0$ & \hspace{0.4 in} First-order & \hspace{0.4 in}  Continuous & \hspace{0.4 in}  First-order \tabularnewline
\hline
Directional dependence of in-plane $B_{c2}$/$\chi_s$  & \hspace{0.4 in} Anisotropic & \hspace{0.4 in}  Isotropic & \hspace{0.4 in} Isotropic \tabularnewline
\hline\hline
\end{tabular}
\label{table:01}
\end{table*}   

{\emph {First-order phase transition at $B_{c2}$ in low temperature regime.}}---To understand the nature of the phase transition at $B_{c2}$ in the low temperature regime, we study how the free energy of a superconducting monolayer WTe$_2$ evolves under $\bm{B}$. Based on the full $H_{BdG}(\bm{k})$ in Eq.\ref{eq:HBdG}, the free energy of the SOPC superconductor as a function of $\Delta$ can be obtained as \cite{NoteX,altland}:
\begin{equation}
f_s=\frac{V|\Delta|^2}{U}-\frac{1}{\beta}\sum_{\bm{k},n}\ln(1+e^{-\beta \epsilon_{\bm{k},n}}),
\end{equation}
where  $\epsilon_{\bm{k},n}$ are the quasi-particle energies of $H_{BdG}(\bm{k})$. With fixed SOPC strength $A_y=1.71$ eV$\cdot$\AA\ and carrier density $n=10\times 10^{12} cm^{-2}$, the evolution of $f_s-f_n$ at $T = 0.1 T_c$ under increasing $B$ is shown in Fig.\ref{fig4}a (note that $f_n \equiv f_s(\Delta = 0)$). Clearly, for $0<B<B_{c2}$, a local minimum in the free energy landscape develops at $\Delta=0$(purple curve)  and eventually becomes the global minimum at $B = B_{c2}$(red curve), where the superconductor-metal transition occurs. Notably, $\Delta$ drops abruptly to zero at $B_c$, which signifies a first-order phase transition. 

The full self-consistent $B-T$ phase diagram from minimizing $f_s-f_n$ is shown in Fig.\ref{fig4}b with the phase boundary at $B_{c2}$ accurately captured for all $T<T_c$. In accord with Fig.\ref{fig4}a, the order parameter drops abruptly to zero at $B_{c2}$ in the low temperature regime. We note that the mechanism of first-order transition in the low temperature limit for superconductors with SOPC is similar to a conventional superconductor, but the phase transition happens much higher than $B_p$ in SOPC superconductors as illustrated in Fig.\ref{fig4}b. In particular, this distinctive first-order transition in the SOPC superconductor WTe$_2$ is very different from the continuous phase transition found in noncentrosymmetric Ising superconductors such as NbSe$_2$ due to a significant $\chi_s$ induced by Ising spin-orbit couplings \cite{Wakatsuki, Sohn2018, Yingming}.

 {\emph {Conclusion and Discussions.}}--- In this work, we identified a new class of centrosymmetric spin-orbit-parity coupled superconductors where SOPC leads to enhancement of in-plane $B_{c2}$ higher than $B_p$. In particular, we explained how the strong parity-mixing due to SOPC near the topologically nontrivial gap edge gives rise to a strongly enhanced $B_{c2}$ in the superconducting topological monolayer WTe$_2$ with low electron carrier density. We further pointed out that the $B_{c2}$ of SOPC superconductors can exhibit an anisotropy in in-plane field directions (but the anisotropy has not yet been observed experimentally). These properties are distinguished from both conventional superconductors and Ising superconductors as summarized in Table \ref{table:01}. 

While we considered an SOPC superconductor in the clean limit, we briefly discuss here the effect of disorder. By including potential fluctuation scattering and spin-orbit scattering effects in the Green function and the vertex correction to $\chi_s$, we show that the $B_{c2}$ is not sensitive to potential fluctuation scattering but a finite $\chi_s$ is induced by spin-orbit scattering, which further enhances the $B_{c2}$ \cite{NoteX}. This explains why a higher $B_{c2}\approx 4B_{p}$ was observed in the more disordered sample \cite{Fatemi}. 

In the main text, we assumed intra-orbital pairing in Eq.\ref{eq:HBdG} belonging to the $A_g$ representation of the $C_{2h}$ point group. Here, we discuss the effect of an inter-orbital singlet pairing: $\hat{\Delta}_1 = \Delta_1 \eta_1 s_x$, which belongs to the $B_u$ representation of $C_{2h}$. First, we show that the $B_u$ phase can be favored only when the band mixing due to SOPC is strong because significant contributions from both parity-odd and parity-even orbitals at the Fermi energy are needed for the pairing to be effective \cite{NoteX}. Interestingly, such an odd-parity pairing leads to a DIII class topological superconductor when the Fermi surface encloses odd number of time-reversal-invariant-momentum(TRIM) points \cite{Fuliang}. In fact, projecting $\hat{\Delta}_1$ to the MCPB basis explicitly reveals that the combination of $\hat{\Delta}_1$ and SOPC results in an effective $p_x \pm ip_y$ pairing \cite{NoteX}.

Unfortunately, superconductivity in monolayer WTe$_2$ sets in when the Fermi surface consists of two disconnected $Q$-pockets away from the TRIM points (Fig.\ref{fig2}). Thus, the system remains topologically trivial. Only by artificially tuning the chemical potential to enclose the $\Gamma$ point, helical Majorana modes can emerge on the edge \cite{NoteX}. Moreover, the effective $p$-wave pairing can result in large $\chi^{yy}_{s}$ and divergent $B^{yy}_{c2}$ which were not observed experimentally \cite{Fatemi, Sajadi}. Thus, we believe that the $B_u$ phase is less likely to be manifested experimentally in WTe$_2$.

{\emph {Note.}}---After presenting the main findings of this work \cite{March_meeting_link}, we noticed that the enhancement of $B_{c2}$ was observed in non-topological centrosymmetric materials without band inversion such as in few-layer stanene and ultrathin PdTe$_2$~\cite{TypeII_wangcong, typeII_Falson,wangjian}. The enhanced $B_{c2}$ in these materials originates mainly from $\bm{k}$-independent atomic spin-orbital coupling, which is very different from the SOPC effect studied in this work.

 {\emph {Acknowledgments.}}---The authors thank Wenyu He, Noah F.Q. Yuan for discussions and Mengli Hu and Junwei Liu for showing us the band structure of 1T'-WTe$_2$ from first-principle calculations. KTL acknowledges the support of the Croucher Foundation and HKRGC through C6025-19G, C6026-16W, 16310219 and16309718.

\clearpage

\onecolumngrid
\begin{center}
	\textbf{\large Supplementary Material for `Spin-orbit-parity coupled superconductivity in topological monolayer WTe$_2$'}\\[.2cm]
Ying-Ming Xie,$^{1,*}$ Benjamin T. Zhou,$^{1,*}$ and K. T. Law$^{1,\dagger}$\\[.1cm]
	{\itshape ${}^1$Department of Physics, Hong Kong University of Science and Technology, Clear Water Bay, Hong Kong, China}\\
	(Dated: \today)\\[1cm]
\end{center}
\setcounter{equation}{0}
\setcounter{section}{0}
\setcounter{figure}{0}
\setcounter{table}{0}
\setcounter{page}{1}
\renewcommand{\theequation}{S\arabic{equation}}
\renewcommand{\thetable}{S\arabic{table}}
\renewcommand{\thesection}{\arabic{section}}
\renewcommand{\thefigure}{S\arabic{figure}}
\renewcommand{\bibnumfmt}[1]{[S#1]}
\renewcommand{\citenumfont}[1]{S#1}
\makeatletter
\section{$\textbf{k}\cdot \textbf{p}$ model of monolayer 1T$'$-WTe$_2$}
  Here, we present detailed derivation of the $\bm{k\cdot p}$ Hamiltonian in Eq.1 of the main text based on the mirror symmetry $M_y$, inversion symmetry $P$ and time reversal symmetry $T$. According to first principle calculations \cite{Qian1256815,Tang2017S,PhysRevX.6.041069,PhysRevB.95.245436,PhysRevB.93.125109,JCsong}, the dominant orbitals near $\Gamma$ point transforms as $p_y$ and $d_{yz}$ orbitals, which have opposite spatial parities and are odd under $M_y$. In the basis $(\ket{p_y,\uparrow},\ket{p_y,\downarrow}, \ket{d_{yz},\uparrow},\ket{d_{yz},\downarrow})$, the symmetry operators are given by: $M_y= -i\sigma_y, P=s_z, T=i\sigma_y K$ \cite{Note}. Using the method of invariant \cite{PhysRevB.82.045122}, we write down a four-band $\bm{k\cdot p}$ model as
\begin{equation}\label{eq:H2}
H_0(\bm{k})=\begin{pmatrix}
\epsilon_{p}(\bm{k})&0&-ivk_x+A_zk_y&-iA_xk_x+A_yk_y\\
0&\epsilon_{p}(\bm{p})&iA_xk_x+A_yk_y&-ivk_x-A_zk_y\\
ivk_x+A_zk_y&-iA_xk_x+A_yk_y&\epsilon_{d}(\bm{k})&0\\
i A_xk_x+A_yk_y&ivk_x-A_zk_y&0&\epsilon_{d}(\bm{k})
\end{pmatrix},
\end{equation}
where $\epsilon_{p}(\bm{k})=-t_{xp}k_x^2-t_{yp}k_y^2-\mu_p$, $\epsilon_{d}(\bm{k})=-t_{xd}k_x^2-t_{yd}k_y^2+t'_{x}k_x^4+t'_{y}k_y^4-\mu_d$.  The effective parameters in Eq.\ref{eq:H2} (listed in Table \ref{table2}) are determined by fitting the \textit{ab initio} band structure\cite{Qian1256815,Tang2017S}. Note that $A_x$, $A_z < A_y$ due to the highly anisotropic crystal symmetry of 1T'-WTe$_2$ and the values of $A_x,A_y,A_z$ in  Table \ref{table2} are  mainly as a reference for the scale, which is sensitive to the gap and hard to be solely determined by the \textit{ab initio} calculation.

\begin{figure}[ht]
	
	\centering
	\includegraphics[width=0.8\linewidth]{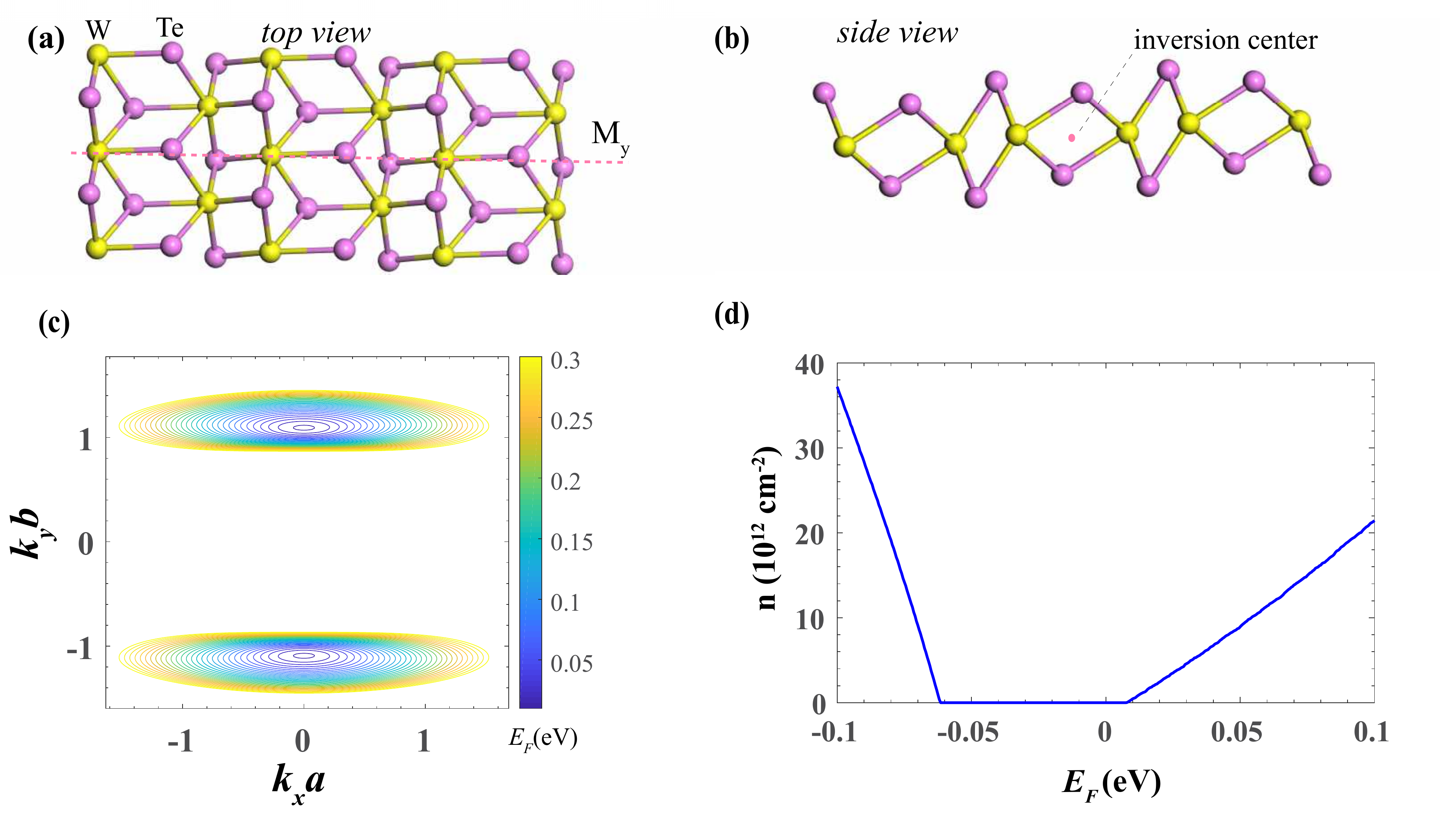}
	\caption{Crystal structure and band structure of monolayer 1T$'$-WTe$_2$. (a) Top view and (b) side view of monolayer WTe$_2$. $W$/$Te$ atoms are depicted in yellow/purple. $M_y$ axis and inversion center are highlighted in pink dashed line and pink dot, respectively.     (c) Fermi circles for Fermi energy $E_F$ in the range $0\sim 0.3$ eV from the model Hamiltonian (\ref{eq:H2}). The separation between adjacent Fermi circles is 0.01 eV. The corrsponding parameters are listed in Table~\ref{table2}.\label{fig:bandcontinuewte2fit} (d) Carrier density $n=\braket{\psi^{\dagger}(\bm{r})\psi(\bm{r})}$ versus $E_F$.  }
%	Note that by chaging the value of $\lambda_y$, the conduction band minimum can be shifted, which will in turn change the correspondence between parameter $E_F$ and $n$. For example, when $\lambda_y=1$, $n=5$ (in unit of $10^{12}$ cm$^{-2}$) $\rightarrow E_F=30$meV, $n=50\rightarrow E_F=200$ meV; when $\lambda_y=2$, $n=10 \rightarrow E_F=175$meV; when $\lambda_y=3$, $n=10\rightarrow E_F=326$ meV. 
	\label{supp_fig1}
\end{figure}

\begin{table}[h]
	\centering
	\caption{Parameters of $\textbf{k}\cdot \textbf{p}$ Hamiltonian (\ref{eq:H2}). The lattice constants are $a=6.31${\AA}, $b=3.49${\AA}.}
	\begin{tabular}{cccccccccccc}
		\hline\hline
		$\mu_p$(eV)&$\mu_d$(eV)& $t_{xp}$($eV$$\cdot${\AA}$^2$)&$t_{yp}$($eV$$\cdot${\AA}$^2$)&$t_{xd}$($eV$$\cdot${\AA}$^2$)&$t_{yd}$($eV$$\cdot${\AA}$^2$)&$t'_{x}$($eV$$\cdot${\AA}$^4$)&$t'_{y}$($eV$$\cdot${\AA}$^4$)&$v$(eV$\cdot${\AA})&$A_{x}$(eV$\cdot${\AA})&$A_{y}$(eV$\cdot${\AA})&$A_{z}$(eV$\cdot${\AA})\\
		\hline
		-1.39&0.062&12.45&18.48 &-2.58&2.68&-7.79&26.65&2.34&0.17&0.57&0.07\\
		\hline
	\end{tabular}	
	\label{table2}
\end{table}

%\begin{equation}
%\chi_t^{ii}=-\frac{1}{2}\mu_B^2\sum_{\bm{k},m, n} \frac{f(E_m(\bm{k}))-f(E_n(\bm{k}))}{E_m(\bm{k})-E_n(\bm{k})+i0^{+}}\braket{n,\bm{k}|\sigma^{i}|m,\bm{k}}\braket{m,\bm{k}|\sigma^{i}|n,\bm{k}}
%\end{equation}

\section{Effective pairing Hamiltonian for SOPC superconductors}\label{section_2}
The normal state electronic property is captured by the $\bm{k\cdot p}$ model (Eq.\ref{eq:H2}). To further describe the superconducting topological monolayer WTe$_2$, we first write down the Bogoliubov–de Gennes Hamiltonian
\begin{equation}\label{eq:FullBdG}
H_{BdG}(\bm{k})=H_0(\bm{k})\eta_3+\frac{1}{2}gu_B\bm{B\cdot\sigma}+\Delta\eta_1,
\end{equation} 
(same as Eq.2 of the main text). Here $\eta_i$ is the Pauli matrix defined in the particle-hole basis. With the full $H_{BdG}(\bm{k})$, the spin susceptibility and free energy of the system can be readily calculated numerically. However, since superconducting pairing forms from states near Fermi energy only, we can further obtain an effective pairing Hamiltonian by projecting $H_0(\bm{k})$ to the conduction bands where the gate-induced superconductivity occurs. As mentioned in the main text, a convenient choice is the manifestly covariant pseudospin basis. In the following, we first derive the corresponding psedudospin basis $\ket{\bm{k},\alpha},\ket{\bm{k},\beta}$ of the doubly degenerate conduction band. Then, we project the Hamiltonian $H_{BdG}(\bm{k})$ into the subspace $\{\psi^{\dagger}_{\bm{k},\alpha},\psi^{\dagger}_{\bm{k},\beta},\psi_{-\bm{k},\beta},-\psi_{-\bm{k},\alpha}\}$, where $\psi^{\dagger}_{\bm{k},\alpha},\psi^{\dagger}_{\bm{k},\beta}$ is the creation operator of $\ket{\bm{k},\alpha},\ket{\bm{k},\beta}$. 

Note that the Hamiltonian (\ref{eq:H2}) can be rewritten as
\begin{equation}
H_0(\bm{k})=\epsilon_0(\bm{k})+\mathcal{M}(\bm{k})s_z+vk_xs_y+A_xk_xs_x\sigma_y+A_yk_ys_x\sigma_x+A_zk_ys_x\sigma_z,
\end{equation}
where
\begin{align}
&\epsilon_0(\bm{k})=t_x^{+}k_x^2+t_y^{+}k_y^2+\frac{1}{2}t'_{x}k_y^4+\frac{1}{2}t'_{y}k_y^4-\mu_0,\\
&\mathcal{M}(\bm{k})=-\delta+t_x^{-}k_x^2+t_y^{-}k_y^2-\frac{1}{2}t'_xk_x^4-\frac{1}{2}t'_{y}k_y^4.
\end{align}    
Here $t_x^{\pm}=-(t_{xp}\pm t_{xd})/2,t_y^{\pm}=-(t_{yp}\pm t_{yd})/2$, $\mu_0=(\mu_d+\mu_p)/2$, $\delta=(\mu_p-\mu_d)/2$. The spin-orbit-parity coupling(SOPC) terms are given by the $A_i (i=x,y,z)$-terms involving the spin $\sigma$-matrices. Due to the presence of SOPC terms, we first diagonalize the spin part with the basis
$\ket{+1}=\cos{\frac{\theta_{\bm{k}}}{2}}\ket{\uparrow}+\sin\frac{\theta_{\bm{k}}}{2}e^{i\phi_{\bm{k}}}\ket{\downarrow})$, $\ket{-1}=-\sin\frac{\theta_{\bm{k}}}{2}e^{-i\phi_{\bm{k}}}\ket{\uparrow}+\cos\frac{\theta_{\bm{k}}}{2}\ket{\downarrow}$. Here $\theta_{\bm{k}}$ and $\phi_{\bm{k}}$ are defined by $(A_yk_y,A_xk_x,A_zk_y)=Ak(\sin\theta_{\bm{k}}\cos\phi_{\bm{k}},\sin\theta_{\bm{k}}\sin\phi_{\bm{k}},\cos\theta_{\bm{k}})$. Then 
\begin{equation}
H_0(\bm{k})=\epsilon_0(\bm{k})+\mathcal{M}(\bm{k})s_z+vk_xs_y+ Aks_x\tau_z.
\end{equation}
$\tau_z$ is the Pauli matrix defined in $(\ket{+1},\ket{-1})$ space. By straightforward diagonalization, the eigenenergy can be obtained as $\epsilon_{\pm}(\bm{k}) = \epsilon_0(\bm{k})\pm\sqrt{\mathcal{M}^2(\bm{k})+v^2k_x^2+A^2k^2}$, and each band has a two-fold degeneracy due to time-reversal and spatial inversion. The corresponding eigenvectors of the conduction band with $\epsilon_{+}(\bm{k})$ are given by
\begin{equation}
\ket{\bm{k},\alpha'}=\frac{1}{N_{\bm{k}}}\begin{pmatrix}
E(\bm{k})+\mathcal{M}(\bm{k})\\
ivk_x+Ak
\end{pmatrix}\otimes\begin{pmatrix}
\cos{\frac{\theta_{\bm{k}}}{2}}\\
\sin\frac{\theta_{\bm{k}}}{2}e^{i\phi_{\bm{k}}}
\end{pmatrix},\  \ket{\bm{k},\beta'}=\frac{1}{N_{\bm{k}}}\begin{pmatrix}
E(\bm{k})+\mathcal{M}(\bm{k})\\
ivk_x-Ak
\end{pmatrix}\otimes \begin{pmatrix}
-\sin\frac{\theta_{\bm{k}}}{2}e^{-i\phi_{\bm{k}}}\\
\cos\frac{\theta_{\bm{k}}}{2}
\end{pmatrix},
\end{equation}
where $E(\bm{k})=\sqrt{\mathcal{M}^2(\bm{k})+v^2k_x^2+A^2k^2}$, the normalization factor $N_{\bm{k}}=\sqrt{(E(\bm{k})+\mathcal{M}(\bm{k}))^2+(v^2k_x^2+A^2k^2)}$.
 We now construct the the pseudospin basis $\ket{\bm{k},\alpha}$, $\ket{\bm{k},\beta}$ with $\ket{\bm{k},\alpha'}, \ket{\bm{k},\beta'}$. Following the general scheme in Ref.~\cite{PhysRevB.87.104505,PhysRevLett.115.026401, PhysRevB.94.180504, 2016arXiv160904152Y}, we first find the representation of spin operators, and construct a new basis formed by linear combinations of $\ket{\bm{k},\alpha'}$, $\ket{\bm{k},\beta'}$ under which the spin-$z$-component operator $\sigma_z$ is diagonal. Then, we choose a proper phase factor such that the new basis vectors transform formally as spins under symmetry operations. Explicitly, the matrix representations of spin in $(\ket{\bm{k},\alpha'},\ket{\bm{k},\beta'})^{T}$ are given by
\begin{align}
&\left\langle \sigma_x \right\rangle =\begin{pmatrix}
\sin\theta_{\bm{k}}\cos\phi_{\bm{k}}&W_{\bm{k}}e^{-i\phi_{\bm{k}}}(\cos\theta_{\bm{k}}\cos\phi_{\bm{k}}+i\sin\phi_{\bm{k}})\\
W^*_{\bm{k}}e^{i\phi_{\bm{k}}}(\cos\theta_{\bm{k}}\cos\phi_{\bm{k}}-i\sin\phi_{\bm{k}})&-\sin\theta_{\bm{k}}\cos\phi_{\bm{k}}
\end{pmatrix},\label{spinx}\\
&\left\langle \sigma_y \right\rangle =\begin{pmatrix}
\sin\theta_{\bm{k}}\sin\phi_{\bm{k}}&-iW_{\bm{k}}e^{-i\phi_{\bm{k}}}(\cos\phi_{\bm{k}}+i\cos\theta_{\bm{k}}\sin\phi_{\bm{k}})\\
iW^*_{\bm{k}}e^{i\phi_{\bm{k}}}(\cos\phi_{\bm{k}}-i\cos\theta_{\bm{k}}\sin\phi_{\bm{k}})&-\sin\theta_{\bm{k}}\sin\phi_{\bm{k}}
\end{pmatrix},\label{spiny}\\
&\left\langle \sigma_z \right\rangle=\begin{pmatrix}
\cos\theta_{\bm{k}}&-W_{\bm{k}}e^{-i\phi_{\bm{k}}}\sin\theta_{\bm{k}}\\
-W^*_{\bm{k}}e^{i\phi_{\bm{k}}}\sin\theta_{\bm{k}}&-\cos\theta_{\bm{k}}\label{spinz}
\end{pmatrix},
\end{align}
where $W_{\bm{k}}=\frac{(E(\bm{k})+\mathcal{M}(\bm{k}))^2-(Ak-ivk_x)^2}{N^2_{\bm{k}}}$. The positive eigenvalues of the above spin matrices are: $(\sqrt{\sin^2\theta_{\bm{k}}\cos^2\phi_{\bm{k}}+|W_{\bm{k}}|^2(\cos^2\theta_{\bm{k}}\cos^2\phi_{\bm{k}}+\sin^2\phi_{\bm{k}})},\sqrt{\sin^2\theta_{\bm{k}}\sin^2\phi_{\bm{k}}+|W_{\bm{k}}|^2(\cos^2\theta_{\bm{k}}\sin^2\phi_{\bm{k}}+\cos^2\phi_{\bm{k}})},\lambda_{\bm{k}})$,
where $\lambda_{\bm{k}}=\sqrt{\cos^2\theta_{\bm{k}}+|W_{\bm{k}}|^2\sin^2\theta_{\bm{k}}}$, $|W_{\bm{k}}|^2=\sqrt{1-A^2k^2/E^2(\bm{k})}$.
By taking proper linear combinations of $\ket{\bm{k},\alpha'},\ket{\bm{k},\beta'}$ \cite{PhysRevB.87.104505,PhysRevLett.115.026401,PhysRevB.94.180504}, the pseudospin basis can be  obtained as
\begin{align}
\ket{\bm{k},\alpha}&=\frac{e^{-i\frac{\alpha_{\bm{k}}}{2}}}{\sqrt{|W_{\bm{k}}|^2\sin^2\theta_{\bm{k}}+(\cos\theta_{\bm{k}}-\lambda_{\bm{k}})^2}}(W_{\bm{k}}\sin\theta_{\bm{k}}\ket{\bm{k},\alpha'}+(\cos\theta_{\bm{k}}-\lambda_{\bm{k}})e^{i\phi_{\bm{k}}}\ket{\bm{k},\beta'})\\
\ket{\bm{k},\beta}&=\frac{e^{i\frac{\alpha_{\bm{k}}}{2}}}{\sqrt{|W_{\bm{k}}|^2\sin^2\theta_{\bm{k}}+(\cos\theta_{\bm{k}}-\lambda_{\bm{k}})^2}}((\lambda_{\bm{k}}-\cos\theta_{\bm{k}})e^{-i\phi_{\bm{k}}}\ket{\bm{k},\alpha'}+W_{\bm{k}}^*\sin\theta_{\bm{k}}\ket{\bm{k},\beta'})
\end{align}
where $e^{i\alpha_{\bm{k}}}=\frac{W(\bm{k})}{|W(\bm{k})|}$. It is straightforward to see $T\ket{\bm{k},\alpha}=\ket{-\bm{k},\beta},T\ket{\bm{k},\beta}=-\ket{-\bm{k},\alpha}, P\ket{\bm{k},\alpha}=\ket{-\bm{k},\alpha},P\ket{\bm{k},\beta}=\ket{-\bm{k},\beta}$ (note that under time-reversal operation $\phi_{\bm{k}}\rightarrow \pi+\phi_{\bm{k}},\theta_{\bm{k}}\rightarrow \pi-\theta_{\bm{k}},\alpha_{\bm{k}}\rightarrow -\alpha_{\bm{k}}$). The representations of $\sigma_x,\sigma_y,\sigma_z$ in the pseudospin basis $\{\ket{\bm{k},\alpha},\ket{\bm{k},\beta}$ are
\begin{align}\label{eq:PauliPseudospin}
\tilde{\sigma}_x(\bm{k})&=\begin{pmatrix}
\frac{1-|W_{\bm{k}}|^2}{2\lambda_{\bm{k}}}\sin(2\theta_{\bm{k}})\cos\phi_{\bm{k}}& |W_{\bm{k}}|e^{-i\phi_{\bm{k}}}(\frac{\cos\phi_{\bm{k}}}{\lambda_{\bm{k}}}+i\sin\phi_{\bm{k}})\\
|W_{\bm{k}}|e^{i\phi_{\bm{k}}}(\frac{\cos\phi_{\bm{k}}}{\lambda_{\bm{k}}}-i\sin\phi_{\bm{k}})&\frac{|W_{\bm{k}}|^2-1}{2\lambda_{\bm{k}}}\sin(2\theta_{\bm{k}})\cos\phi_{\bm{k}}
\end{pmatrix}\nonumber\\
&=|W_{\bm{k}}|(\frac{\cos^2\phi_{\bm{k}}}{\lambda_{\bm{k}}}+\sin^2\phi_{\bm{k}})\rho_1+|W_{\bm{k}}|\sin\phi_{\bm{k}}\cos\phi_{\bm{k}}(\frac{1}{\lambda_{\bm{k}}}-1)\rho_2+\frac{1-|W_{\bm{k}}|^2}{2\lambda_{\bm{k}}}\sin(2\theta_{\bm{k}})\cos\phi_{\bm{k}}\rho_3\\\nonumber
\tilde{\sigma}_y(\bm{k})&=\begin{pmatrix}
\frac{1-|W_{\bm{k}}|^2}{2\lambda_{\bm{k}}}\sin(2\theta_{\bm{k}})\sin\phi_{\bm{k}}& |W_{\bm{k}}|e^{-i\phi_{\bm{k}}}(\frac{\sin\phi_{\bm{k}}}{\lambda_{\bm{k}}}-i\cos\phi_{\bm{k}})\\
|W_{\bm{k}}|e^{i\phi_{\bm{k}}}(\frac{\sin\phi_{\bm{k}}}{\lambda_{\bm{k}}}+i\cos\phi_{\bm{k}})&\frac{|W_{\bm{k}}|^2-1}{2\lambda_{\bm{k}}}\sin(2\theta_{\bm{k}})\sin\phi_{\bm{k}}
\end{pmatrix}\nonumber\\
&=|W_{\bm{k}}|\sin\phi_{\bm{k}}\cos\phi_{\bm{k}}(\frac{1}{\lambda_{\bm{k}}}-1)\rho_1+|W_{\bm{k}}|(\frac{\sin^2\phi_{\bm{k}}}{\lambda_{\bm{k}}}+\cos^2\phi_{\bm{k}})\rho_2+\frac{1-|W_{\bm{k}}|^2}{2\lambda_{\bm{k}}}\sin(2\theta_{\bm{k}})\sin\phi_{\bm{k}}\rho_3\nonumber\\
\tilde{\sigma}_z(\bm{k})&=\lambda_{\bm{k}}\begin{pmatrix}
1&0\\0&-1
\end{pmatrix}=\lambda_{\bm{k}}\rho_3
\end{align}
For notational convenience, we define $\tilde{\sigma}_{i}(\bm{k})=\sum_ja_{ij}(\bm{k})\rho_j$. $a_{ij}(\bm{k})$ captures the effect of SOPC on the spin properties, $\rho_j$ is the Pauli matrix defined in the pseudospin basis. It can be verified that under all symmetry operations, $\tilde{\sigma}_{i}(\bm{k})$ has the same transformation rules as spins. By projecting the full BdG Hamiltonian to the pseudospin basis, the pairing Hamiltonian is 
\begin{equation}
H_s=\Delta\sum_{\bm{k}}c_{\bm{k}p,\uparrow}^{\dagger}c_{-\bm{k}p,\downarrow}^{\dagger}+c_{\bm{k}d,\uparrow}^{\dagger}c_{-\bm{k}d,\downarrow}^{\dagger}+h.c.\approx\Delta\sum_{\bm{k}}\psi^{\dagger}_{\bm{k},\alpha}\psi^{\dagger}_{-\bm{k},\beta}+h.c..
\end{equation}
Note that the form of $s$-wave pairing is preserved, \textit{i.e.}, pseudospin-up and pseudospin-down states with opposite momentum are paired. This leads to the final form (as in Eq.3 of the main text) of the effective pairing Hamiltonian in the Nambu pseudospin basis $\Psi_{\bm{k}}^{\dagger}=(\psi^{\dagger}_{\bm{k},\alpha},\psi^{\dagger}_{\bm{k},\beta},\psi_{-\bm{k},\beta},-\psi_{-\bm{k},\alpha})$:
\begin{equation}
H_{\text{eff}}=\sum_{\bm{k},l,l'}\psi^{\dagger}_{\bm{k},l}(\xi_{+}(\bm{k})\delta_{l,l'}+\frac{1}{2}g_su_B\bm{B\cdot}\bm{\tilde{\sigma}}_{l,l'}(\bm{k}))\psi_{\bm{k},l'}+\Delta\sum_{\bm{k}}\psi^{\dagger}_{\bm{k},\alpha}\psi^{\dagger}_{-\bm{k},\beta}+h.c.,
\end{equation}
where $l$ labels $\alpha,\beta$, $\xi_{\pm}(\bm{k})=\epsilon_0(\bm{k})+E(\bm{k})$. In the following, we neglect the $+$ index, \textit{i.e.}, $\xi_{\bm{k}}\equiv \xi_{+}(\bm{k})$. 

\section{Pauli spin susceptibility and renormalization factor $\gamma_i$}

In general, the Pauli spin susceptibility with mean-field order parameter $\Delta$ is given by

\begin{equation}
\chi^{ij}_{s}=-\frac{1}{2}u_B^2k_BT\sum_{\bm{k}}\sum_{\omega_n}\text{Tr}[\tilde{\sigma_i}\mathcal{G}^{0}(\bm{k},i\omega_n)\tilde{\sigma}_j\mathcal{G}^{0}(\bm{k},i\omega_n)].\label{spin_sus_0}
\end{equation}
Here $\mathcal{G}^{0}(\bm{k},i\omega_n)=(i\omega_n-\xi_{\bm{k}}\eta_3-\Delta\eta_1)^{-1}=-\frac{i\omega_n+\xi_{\bm{k}}\eta_3+\Delta\eta_1}{\omega_n^2+\xi_{\bm{k}}^2+\Delta^2}$ is the Nambu-Gor'kov Green's function. The factor $1/2$ results from the particle-hole redundancy of Nambu basis.
Eq.\ref{spin_sus_0} is equivalent to the spin susceptibility formula given in Ref.~\cite{Sigrist_susceptibility,abrikosov1962spin}. By tracing out the psedospin and particle-hole indices, we obtain
 \begin{align}
 \chi_s^{ii}=-u^2_{B}k_BT\sum_{\bm{k}}\sum_{\omega_n}\gamma_{i}(\bm{k})\frac{-\omega_n^2+\xi_{\bm{k}}^2+\Delta^2}{(\omega_n^2+\Delta^2+\xi^2_{\bm{k}})^2}\label{spin_sus_3}
 \end{align}
where $\gamma_{i}(\bm{k})=2\sum_{j}a^2_{ij}(\bm{k})$ is the renormalization factor due to SOPCs, which are given explicitly by
\begin{align}\label{gamma1}
&\gamma_{x}(\bm{k})=\frac{4|W_{\bm{k}}|^2+(1-|W_{\bm{k}}|^2)^2\sin^22\theta_{\bm{k}}}{2(\cos^2\theta_{\bm{k}}+|W_{\bm{k}}|^2\sin^2\theta_{\bm{k}})}\cos^2\phi_{\bm{k}}+2|W_{\bm{k}}|^2\sin^2\phi_{\bm{k}}\\
&\gamma_{y}(\bm{k})=\frac{4|W_{\bm{k}}|^2+(1-|W_{\bm{k}}|^2)^2\sin^22\theta_{\bm{k}}}{2(\cos^2\theta_{\bm{k}}+|W_{\bm{k}}|^2\sin^2\theta_{\bm{k}})}\sin^2\phi_{\bm{k}}+2|W_{\bm{k}}|^2\cos^2\phi_{\bm{k}}\label{gamma2}\\
&\gamma_{z}(\bm{k})=2(\cos^2\theta_{\bm{k}}+|W_{\bm{k}}|^2\sin^2\theta_{\bm{k}})\label{gamma3}.
\end{align}
By summing over the Matsubara frequencies in Eq.\ref{spin_sus_3} first, the form of spin susceptibility can be further simplified to 
\begin{align}
\chi_{s}^{ii}=\frac{1}{2}u_{B}^2\beta\sum_{\bm{k}}\gamma_{i}(\bm{k})\frac{1}{1+\cosh(\beta\sqrt{\xi^2_{\bm{k}}+\Delta^2})}. 
\end{align}
Note that at zero temperature, the residue spin susceptibility $\chi^{ii}_s$ vanishes since $\cosh(\beta\sqrt{\xi^2_{\bm{k}}+\Delta^2})\to \infty$. By taking $\Delta\rightarrow0$, the normal-state spin susceptibility is recovered:
\begin{equation}
\chi_{n}^{ii}=\frac{1}{2}u_{B}^2\beta\sum_{\bm{k}}\gamma_{i}(\bm{k})\frac{1}{1+\cosh(\beta\xi_{\bm{k}})}=u_B^2N(E_F)\braket{\gamma_i(E_F)}\label{spinsus1}.
\end{equation}
Here $\braket{\gamma_i(E_F)}=\int d^2 \bm{k} \gamma_i(\bm{k})\delta(\xi_{\bm{k}}-E_F)/\int d^2 \bm{k} \delta(\xi_{\bm{k}}-E_F)$ is the average value of $\gamma_i$ over the Fermi surface. Obviously, in the zero temperature limit, the normal-state spin susceptibility is controlled by $\gamma_i(\bm{k})$, which can take a value within $[0, 2]$. To see how $\gamma_i(\bm{k})$ is affected by SOPCs, we note that if SOPC is absent, \textit{i.e.}, $Ak=0$, then $|W_{\bm{k}}|=1$ and $\gamma_i=2$. As we discussed in the main text, in this case the in-plane $B_{c2}$ reduces to $B_p$. Upon increasing the SOPC strength, $|W_{\bm{k}}|$ is reduced, which reduces the value of $\gamma_i(\bm{k})$ and results in $B_{c2}>B_p$. 

%At zero temperature, $B_{c2}$ can be estimated as
%\begin{equation}
%B_{c2}=\sqrt{\frac{N(E_F)\Delta^2}{\chi_n-\chi_s}}=\sqrt{\frac{N(E_F)\Delta^2}{\chi_0}}\sqrt{\frac{\chi_0}{\chi_n-\chi_s}}=\frac{\Delta}{\sqrt{2}u_B}\sqrt{\frac{\chi_0}{\chi_n-\chi_s}}=B_{p}\sqrt{\frac{\chi_0}{\chi_n-\chi_s}},
%\end{equation}  
%where $\chi_0=2N(E_F)u_B^2$ is the spin susceptibility of usual electron gas.
%
We note that there is another equivalent form of spin susceptibility obtained by performing the momentum integral first for Eq.\ref{spin_sus_3}:
\begin{equation}\label{clean_sus} 
\chi_s^{ii}/\chi_n^{ii}=1-\pi k_BT\sum_{\omega_n}\frac{\Delta^2}{(\Delta^2+\omega_n^2)^{3/2}},
\end{equation}
where $\chi_n^{ii}$ is the reduced normal spin susceptibility. This form would provide a more straightforward way to understand disorder effects on the enhancement of $B_{c2}$ as we shall discuss in details in Section VI.

\section{$B_{c2}$ from the linearized gap equation}
Here, we present details of the linearized gap equation we used to obtain the enhancement of $B_{c2}$ shown in Fig.3 of the main text. Given the pairing Hamiltonian 
\begin{equation}
H=\sum_{\bm{k},l,l'}\psi^{\dagger}_{\bm{k},l}(\xi_{\bm{k}}\delta_{l,l'}+u_B\bm{B\cdot\tilde{\sigma}})\psi_{\bm{k},l'}-\frac{U}{2V}\sum_{\bm{k},\bm{k'}}\psi^{\dagger}_{\bm{k},\alpha}\psi^{\dagger}_{-\bm{k},\beta}\psi_{-\bm{k'},\beta}\psi_{\bm{k'},\alpha},
\end{equation}
the corresponding linearized gap equation is given by
\begin{equation}
\frac{2}{U/V}=k_BT\sum_{\bm{\bm{k}}}\sum_{n}\text{Tr}[G^{(0)}(\bm{k},i\omega_n)\rho_yG^{(0)T}(-\bm{k},-i\omega_n)\rho_y].
\end{equation}

Upon further simplifications and Matsubara sum, we have
\begin{eqnarray}\label{lineargap}
\frac{1}{U/V}&=&k_BT\sum_{\bm{k}}\sum_{n}\frac{(i\omega_n-\xi_{\bm{k}})(-i\omega_n-\xi_{\bm{k}})-u_B^2\sum_j(\sum_ia_{ij}(\bm{k})B_i)^2}{((i\omega_n-\xi_{\bm{k}})^2-u_B^2\sum_j(\sum_ia_{ij}(\bm{k})B_i)^2)((-i\omega_n-\xi_{\bm{k}})^2-u_B^2\sum_j(\sum_ia_{ij}(\bm{k})B_i)^2)},\\
\frac{1}{U/V}&=&k_BT\sum_{\bm{k}}\frac{\sinh\beta\xi}{2\xi(\cosh\beta\xi+\cosh(\beta u_BB_{eff}(\bm{k})))},
\end{eqnarray}
where $B_{eff}=\sqrt{\sum_i(a_{ij}(\bm{k})B_i)^2}$. For magnetic field along the $i$-direction, $B_{eff}=B_i\sqrt{\gamma_{\bm{k}}/2}$. In the absence of magnetic fields,
\begin{equation}\label{zerofieldgap}
\frac{1}{U/V}=k_BT_{c}\sum_{\bm{k}} \frac{\sinh \beta\xi}{2\xi(\cosh\beta\xi+1)}=k_BT_cN(E_F)\int_{-\hbar\omega_D}^{\hbar\omega_D}  d\xi\frac{\tanh \beta \xi/2}{2\xi}=N(E_F)\ln(\frac{2e^{\gamma}\hbar\omega_D}{\pi k_BT_c}),
\end{equation}
Here, $\gamma$ is the Euler constant, $T_{c}$ is the zero field critical temperature. Substituting the expression of $\frac{1}{U/V}$ in Eq.\ref{zerofieldgap} into Eq.\ref{lineargap}, we get
\begin{equation}
\ln(\frac{T}{T_{c}})=\int_{-\infty}^{+\infty} d\xi \int _0^{2\pi}\frac{d\varphi}{2\pi}\frac{\sinh\beta\xi}{2\xi}(\frac{1}{\cosh\beta\xi+\cosh(\beta u_BB_{eff}(E_F,\varphi))}-\frac{1}{\cosh\beta\xi+1})
\end{equation}
Due to the complicated $a_{ij}(\bm{k})$ coefficients, the linearized gap equation was solved numerically by transforming the energy integral into a summation over momentum.

\section{Derivation of superconducting free energy}

As we discussed in the main text, the scheme of linearized gap equation fails to capture the first-order phase transition at $B_{c2}$ for the centrosymmetric spin-orbit-parity coupled(SOPC) superconductor WTe$_2$. As the in-plane field increases and approaches the superconductor-metal phase boundary, the superconducting gap and the value of $B_{c2}$ need to be determined self-consistently by the minimum of the superconducting free energy $f_s$ of the system. Here, we present a detailed derivation of the expression of $f_s$ in the main text, which allows us to obtain the evolution of $f_s$ under magnetic fields and the full superconducting phase diagram shown in Fig.4 of the main text.

In general, the partition function of a system involving two-body interactions can be written as:
\begin{equation}
Z=\int D[\psi(\bm{r},\tau),\bar{\psi}(\bm{r},\tau)]\exp\{-S[\psi(\bm{r},\tau),\bar{\psi}(\bm{r},\tau)]\},
\end{equation}
where the action is given by
\begin{equation}
S[\psi,\bar{\psi}]=\int d\tau\int d\bm{r} \sum_{\sigma}\bar{\psi}(\bm{r},\tau)\partial_{\tau}\psi(\bm{r},\tau)+\sum_{\sigma\sigma'}\bar{\psi}_{\sigma}(\bm{r},\tau)H_0(\bm{r},\tau)\psi_{\sigma'}(\bm{r},\tau)-g\sum_{\sigma\sigma'}\bar{\psi}_{\sigma}(\bm{r},\tau)\bar{\psi}_{\sigma'}(\bm{r},\tau)\psi_{\sigma'}(\bm{r},\tau)\psi_{\sigma}(\bm{r},\tau).
\end{equation}

By introducing an auxiliary bosonic field, the interaction term can be reformulated via the Hubbard-Stratonovich transformation:
\begin{equation}
\exp(g\int d\tau\int d\bm{r}\bar{\psi}_{\uparrow}\bar{\psi}_{\downarrow}\psi_{\downarrow}\psi_{\uparrow})=\int D[\bar{\Delta},\Delta]\exp(-\int d\tau\int d\bm{r}[\frac{1}{g}|\Delta|^2-\Delta\bar{\psi}_{\uparrow}\bar{\psi}_{\downarrow}-\bar{\Delta}\psi_{\downarrow}\psi_{\uparrow}]),
\end{equation}
Then, the action becomes
\begin{equation}
Z=\int D[\bar{\psi}(\bm{r},\tau),\psi(\bm{r},\tau)]\int D[\bar{\Delta},\Delta]\exp(-S).
\end{equation}
Here
\begin{equation}
S=\frac{1}{2}\int d\tau \int d\bm{r} \bar{\Phi}G^{-1}\Phi+\frac{1}{g}|\Delta|^2,
\end{equation}
where $\Phi=(\bar{\psi}_{\uparrow},\bar{\psi}_{\downarrow},\psi_{\uparrow},\psi_{\downarrow})$ and
\begin{equation}
G^{-1}=\begin{pmatrix}
\partial_{\tau}+H_0&\Delta i\sigma_y\\(\Delta i\sigma_y)^{\dagger}&\partial_{\tau}-H_0^{*}
\end{pmatrix}.
\end{equation}
Integrate out the Grassman field $\psi(\bm{r},\tau)$, we have
\begin{equation}
Z=\int D[\bar{\Delta},\Delta]\exp(-S_{eff}),
\end{equation}
where $S_{eff}=\int d\tau\int d\bm{r}\frac{1}{g}|\Delta|^2+\ln\text{Det}G^{-1}$. Within the mean-field approximation, $\Delta$ is assumed to be uniform in space and time. This reduces the mean-field free energy to the form
\begin{equation}
f_s=\frac{1}{\beta}\ln(Z)=\frac{1}{\beta}S_{eff}=\frac{V}{g}|\Delta|^2-\frac{1}{\beta}\ln\text{Det}G^{-1}=\frac{V}{g}|\Delta|^2-\frac{1}{\beta}\sum_{\bm{k},n}\ln(1+e^{-\beta \epsilon_{\bm{k},n}}).
\end{equation}
Here, $V$ is the volume of system. The quasi-particle energies $\epsilon_{\bm{k},n}$ are calculated from the full Bogoliubov–de Gennes Hamiltonian $H_{BdG}=H_0(\bm{k})\eta_3+\frac{1}{2}g_su_B\bm{B\cdot\sigma}+\Delta\eta_1$.

%\section{Further discussion}
%
%The spin-orbital scattering theory in this inversion symmetrical system.
%
%Construct a model that has both inversion and time reversal symmetry but gives finite superconducting spin susceptibility.
%
%why for non-centrosymmetrical superconductor we don't pay too much attention to this effect?
%
%
%emphasis low gating region but high gating does not
%How the disorder affects our picture?
%
%whether it is general for a four band model with time reversal and inversion symmetry  after band inversion, the spin susceptibility in the band bottom is reduced?
%
%The misunderstanding that if there is SOC, the spin susceptibility is reduced.  (see our model in high Fermi energy region, noncentrosymmetical case)
%
%$E_F\tau$ is small means the quantum correction is strong
%
%change FIg into Fig s2
\section{Spin susceptibility with non-magnetic impurity scattering}

\begin{figure}
	\centering
	\includegraphics[width=0.5\linewidth]{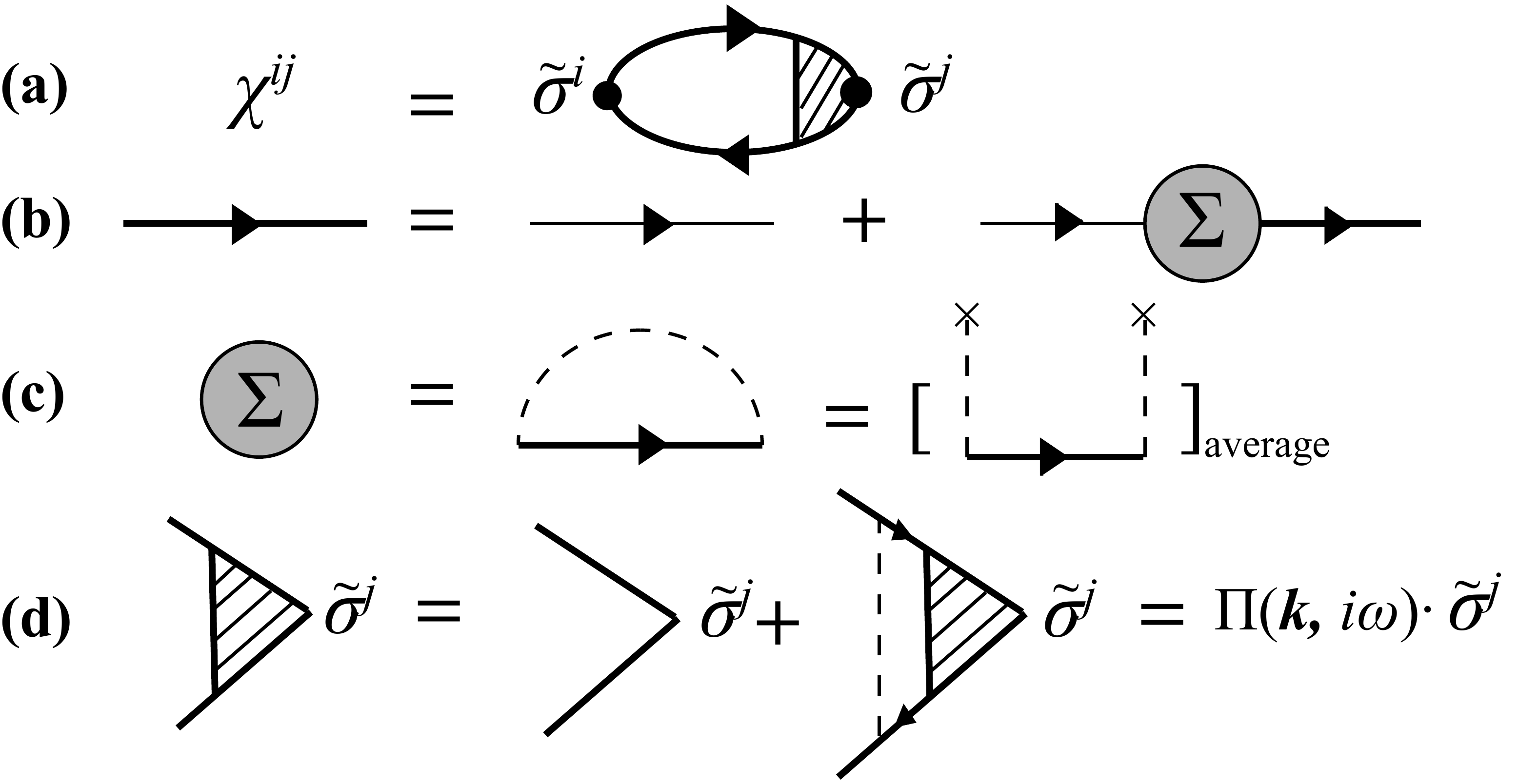}
	\caption{Diagrammatic representation of (a) disorder-averaged spin susceptibility, (b) Dyson equation for self-energy correction, (c)  self-energy in self-consistent Born approximation, (d)  integral equation for spin vertex correction. The impurity potential is $
		U_{im}(\bm{k}-\bm{k'})=U_1(\bm{k}-\bm{k'})+U_2(\bm{k}-\bm{k'})i(\hat{\bm{k}}\times\hat{\bm{k'}})\cdot\bm{\sigma} $ \cite{RevModPhys.78.373}. The first term describes scattering from scalar potential fluctuations, and the second term describes the spin-orbit scattering. }
	\label{fig:spinsusdisorder}
\end{figure}

We discussed briefly in the main text that the enhancement of $B_{c2}$ in SOPC superconductor is not affected in a qualitative way by disorder. Here we present detailed analysis of disorder effects on the SOPC superconductor WTe$_2$. Including both local potential fluctuation and spin-orbit scattering, the non-magnetic impurity potential can be written as \cite{KLB}
\begin{align}
U_{im}(\bm{k}-\bm{k'})&=U_1(\bm{k}-\bm{k'})\eta_3+U_2(\bm{k}-\bm{k'})i(\hat{\bm{k}}\times\hat{\bm{k'}})\cdot\bm{\sigma} \eta_3\\
&=U_1(\bm{k}-\bm{k'})\eta_3+U_2(\bm{k}-\bm{k'})i\lambda_{\bm{k}}\rho_3\eta_3 (\hat{\bm{k}}\times\hat{\bm{k'}})\cdot \hat{z}
\end{align}
The diagrammatic calculation process to obtain the disorder-averaged spin suscepbility under $U_{im}(\bm{k}-\bm{k'})$ is shown in Fig.\ref{fig:spinsusdisorder}: following similar procedures in previous works \cite{abrikosov1962spin,abrikosov1959theory,KLB,dora2002impurity,RevModPhys.78.373}, we first calculate the self-energy correction with the standard Born approximation. Then, we calculate the ladder diagram for the spin vertex correction, and finally obtain the disorder-averaged spin susceptibility as:
\begin{equation}
\overline{\chi_s^{ij}}=-\frac{1}{2}u_B^2k_BT\sum_{\bm{k}}\sum_{\omega_n}\text{Tr}[\tilde{\sigma}_i\mathcal{G}(\bm{k},i\omega_n)\Pi(\bm{k},i\omega_n)\cdot\tilde{\sigma}_j\mathcal{G}(\bm{k},i\omega_n)].\label{spin_sus_imp}
\end{equation} 
Here $\mathcal{G}(\bm{k},i\omega_n)=(i\omega_n-\xi_{\bm{k}}\eta_3-\Delta\eta_1-\Sigma(\bm{k},i\omega_n))^{-1}$ is the Nam-Gor'kov Green's function including the self-energy correction due to disorder. The self-energy $\Sigma(\bm{k},i\omega_n)$ is given by the self-consistent equation
\begin{equation}
\Sigma(\bm{k},i\omega_n)=\int_{\bm{k'}}U_{im}(\bm{k}-\bm{k'}) \mathcal{G}(\bm{k'},i\omega_n)U_{im}(\bm{k'}-\bm{k}),
\end{equation} 
where $\int_{\bm{k}}\equiv \int \frac{d^2\bm{k}}{(2\pi)^2}$. Within the Born approximation, the equation can be solved as
\begin{equation}
\Sigma(\bm{k},i\omega_n)=-\frac{i\omega_n}{\tau\sqrt{\omega_n^2+\Delta^2}}+\frac{\Delta}{\tau\sqrt{\omega_n^2+\Delta^2}}\eta_1,
\end{equation}
where $1/\tau=1/\tau_0+1/\tau_{so}$ with
\begin{align}
\frac{1}{\tau_0}&=\pi N(E_F)\int d^2 \bm{k'} \delta(\xi_{\bm{k}'}-E_F) |U_1(\bm{k}-\bm{k'})|^2,\\
\frac{1}{\tau_{so}}&=\pi N(E_F) \int d^2 \bm{k'} \delta(\xi_{\bm{k}'}-E_F) \lambda_{\bm{k}}^2|U_2(\bm{k}-\bm{k'})|^2\sin^2\varphi_{\bm{k'}}.
\end{align}
Here, $\tau$ is the total scattering time, $\tau_{0}$ is the momentum relaxation time, $\tau_{so}$ is the spin-orbit scattering time. Similar to previous works \cite{abrikosov1962spin,abrikosov1959theory,dora2002impurity,RevModPhys.78.373},  we consider the leading order $s$-wave scattering channel only, thus $\tau$ can be treated as $\bm{k}$-independent. Then $\mathcal{G}(\bm{k},i\omega_n)$ can be rewritten as $\mathcal{G}(\bm{k},i\omega_n)=(i\tilde{\omega}_n-\xi_{\bm{k}}\eta_3-\tilde{\Delta}\eta_1)^{-1}$, where 
\begin{equation}
\tilde{\omega}_n=\omega_n+\frac{\omega_n}{\tau\sqrt{\omega_n^2+\Delta^2}},\  \tilde{\Delta}=\Delta+\frac{\Delta}{\tau\sqrt{\omega_n^2+\Delta^2}}\label{rp}.
\end{equation}

Now, we use $\mathcal{G}(\bm{k},i\omega)$ to calculate the spin vertex corrections. The recursive integral equation for vertex correction, as depicted by the Feynman diagram Fig.\ref{fig:spinsusdisorder}d, is given by
\begin{equation}
\Pi(\bm{k},i\omega_n)\cdot\tilde{\sigma}_j=\tilde{\sigma}_j+\int_{\bm{k'}}U_{im}(\bm{k}-\bm{k'})\mathcal{G}(\bm{k'},i\omega_n)\Pi(\bm{k'},i\omega_n)\cdot \tilde{\sigma}_j\mathcal{G}(\bm{k'},i\omega_n)U_{im}(\bm{k'}-\bm{k})\label{verrtex_correction}
\end{equation}
Here, $\Pi(\bm{k},i\omega)\cdot \tilde{\sigma}_j$ is the spin vertex function, which can be decomposed as 
\begin{equation}
\Pi(\bm{k},i\omega)\cdot \tilde{\sigma}_j\equiv\sum_m \Pi_{m}(\bm{k},i\omega)\braket{a_{jm}(E_F)}\rho_m,
\end{equation}
where $a_{im}(E_F)=\int d^2 \bm{k} a_{im}(\bm{k})\delta(\xi_{\bm{k}}-E_F)$ and $a_{jm}(\bm{k})$ is given in Sec.~\ref{section_2}. A self-consistent ansatz of $\Pi_{m}$ for the integral equation above has the form
\begin{equation}
\Pi_{m}=\lambda_m^{0}+\lambda^{1}_m\eta_1.
\end{equation}
Substitute it into Eq.\ref{verrtex_correction}, we have
\begin{align}
&\lambda_m^0=1+\frac{\tilde{\Delta}^2}{\tau_m(\tilde{\Delta}^2+\tilde{\omega}^2)^{3/2}}\lambda_m^0+\frac{i\tilde{\omega}\tilde{\Delta}}{\tau_m(\tilde{\Delta}^2+\tilde{\omega}^2)^{3/2}}\lambda_m^1,\\
&\lambda^1_m=-\frac{i\tilde{\omega}\tilde{\Delta}}{\tau_m(\tilde{\Delta}^2+\tilde{\omega}^2)^{3/2}}\lambda_m^0+\frac{\tilde{\omega}^2}{\tau_m(\tilde{\Delta}^2+\tilde{\omega}^2)^{3/2}}\lambda_m^1,
\end{align}
where $1/\tau_m=1/\tau_0-1/\tau_{so}$ for $m=1,2$ and $1/\tau_m=1/\tau_0+1/\tau_{so}=1/\tau$ for $m=3$. Then we obtain
\begin{equation}
\lambda_m^0=\frac{\tilde{\omega}^2}{\tilde{\Delta}^2+\tilde{\omega}^2}+\frac{\tilde{\Delta}^2\tau_m}{(-\sqrt{\tilde{\Delta}^2+\tilde{\omega}^2}+\tau_m(\tilde{\Delta}^2+\tilde{\omega}^2))},\ \lambda_m^1=\frac{-i\tilde{\Delta}\tilde{\omega}}{(\tilde{\Delta}^2+\tilde{\omega}^2)(-1+\tau_m\sqrt{\tilde{\Delta}^2+\tilde{\omega}^2})}.
\end{equation}
Simplify them with Eq.\ref{rp}, we get
\begin{equation}
\lambda_m^0=1+\frac{\Delta^2}{\Delta^2+\omega^2}\frac{1}{\tau_m\sqrt{\Delta^2+\omega^2}+(\tau_m/\tau-1)},\ \lambda_m^1=-\frac{i\Delta\omega}{\Delta^2+\omega^2}\frac{1}{\tau_m\sqrt{\Delta^2+\omega^2}+(\tau_m/\tau-1)}.
\end{equation}
Before proceeding to the final result, we discuss more about the vertex correction coefficients  $\lambda_m^0$ and $\lambda_m^1$ here. When $\tau_{so}\rightarrow \infty$, namely, in the absence of spin-orbit scattering, we find $\lambda^0_m=1+\frac{\Delta^2}{\tau_0(\Delta^2+\omega^2)^{3/2}}$, $\lambda^1_m=-i\frac{\Delta\omega}{\tau_0(\Delta^2+\omega^2)^{3/2}}$. The vertex correction function in this case is  $\Pi_{m}(i\omega)=(1-\frac{\partial \Sigma(i\omega)}{\partial i\omega}$), which is exactly the Ward's identity. We have this identity here because without spin-orbit scattering, the vertex behaves as a scaler and spin is a conserve quantity.

After taking both the self-energy and vertex corrections, we can evaluate the disorder-averaged spin susceptibility $\overline{\chi_s^{ii}}$ from Eq.\ref{spin_sus_imp}:
 \begin{equation}
\overline{\chi_s^{ii}}/\chi_n^{ii}=1-\pi k_BT\sum_{\omega_n} \frac{\Delta^2}{(\omega_n^2+\Delta^2)^{\frac{3}{2}}}I^i_{E_F}(\omega_n,\Delta,\tau_0,\tau_{so}),
\end{equation}
 where $\chi_n^{ii}$ is the reduced Pauli spin susceptibility and 
 \begin{equation}
 I^{i}_{E_F}=\sum_m\frac{2\braket{a^2_{im}(E_F)}}{\braket{\gamma_i(E_F)}}\frac{1+\frac{1}{\tau_m\sqrt{\omega_n^2+\Delta^2}+\tau_m/\tau-1}}{1+\frac{1}{\tau\sqrt{\omega_n^2+\Delta^2}}}.\label{factor_disorder}
 \end{equation}
 
   \begin{figure}[h]
 	\centering
 	\includegraphics[width=0.5\linewidth]{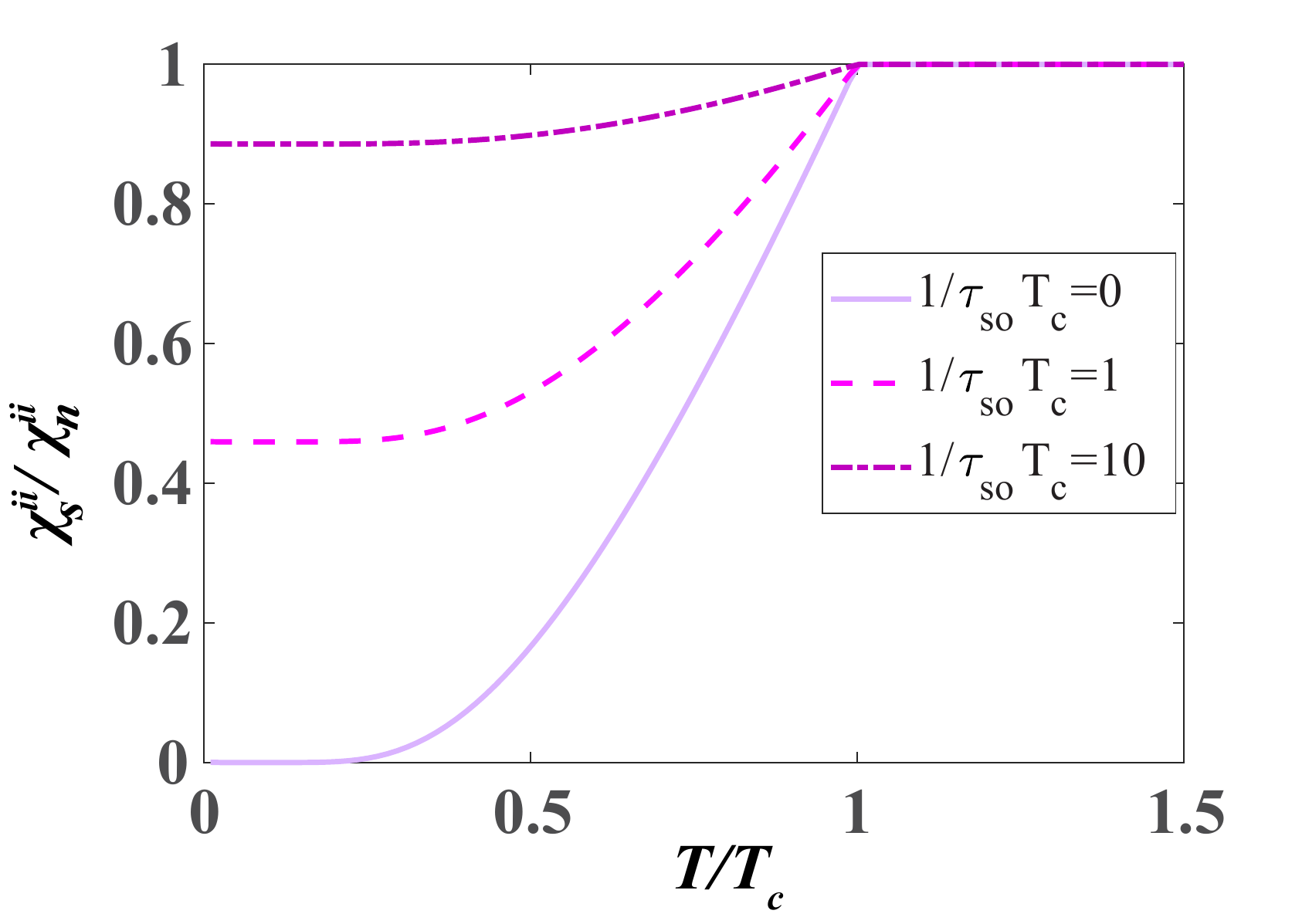}
 	\caption{Plot of $\chi^{ii}_s/\chi^{ii}_n$ versus $T/T_c$ in Eq.\ref{spin_sus_spinorbit} for $i=x,y$. The strength of spin-orbit scattering is characterized by the dimensionless parameter $1/\tau_{so}T_c$. Evidently, the appearance of spin-orbit coupling generates finite residue spin susceptibility that can enhance the upper critical field.}
 	\label{fig:spinorbitsus}
 \end{figure}
 
Comparing with $\chi^{ii}_s$ in the clean case, we have an extra factor $ I^{i}_{E_F}$ here that encodes the information of impurity scattering. When the spin-orbit scattering is absent, namely in the limit $\tau_{so}\rightarrow \infty$, we have $\tau_m = \tau_0 = \tau$, thus $I^{i}_{E_F} = 1$ and we have
\begin{equation}
\overline{\chi_s^{ii}}/\chi_n^{ii}=1-\pi k_BT\sum_{\omega_n}\frac{\Delta^2}{(\Delta^2+\omega_n^2)^{3/2}} = \chi_s^{ii}/\chi_n^{ii}.
\end{equation}
Thus, without spin-orbit scattering, $\overline{\chi_s} = \chi_s$ as shown in Eq.\ref{clean_sus} and the disorder-averaged spin susceptibility is unchanged. This can be seen directly from the ladder diagram: when the scalar Ward's identity is preserved, the self-energy correction cancels the vertex correction. This shows that the $B_{c2}$ in SOPC superconductors is insensitive to the potential fluctuations induced by impurities. 

With finite spin-orbit scattering, namely $\tau_{so}^{-1} \neq 0$,
\begin{equation}\label{spin_sus_spinorbit}
 I^{i}_{E_F}(\omega_n,\Delta,\tau_0,\tau_{so})=\frac{2\braket{a_{i3}(E_F)}}{\braket{\gamma_i(E_F)}}+\sum_{m=1}^{2}\frac{2\braket{a^2_{im}(E_F)}}{\braket{\gamma_i(E_F)}}\frac{1-\frac{1}{\tau_{so}\sqrt{\omega_n^2+\Delta^2}+2}}{1+\frac{1}{\tau_{so}\sqrt{\omega_n^2+\Delta^2}}}.
\end{equation}
The coefficients $\braket{a_{im}(E_F)}$ and $\braket{\gamma_i(E_F)}$ capture the effect of SOPC on spin-orbit scattering. Plots of residue $\chi_s^{ii} (i=x,y)$ at different spin-orbit scattering strengths are shown in Fig.\ref{fig:spinorbitsus}. Clearly, the presence of sufficiently strong spin-orbit scattering with $1/\tau_{so} \sim T_c \sim 0.1$ meV can give rise to a residue spin susceptibility to enhance the $B_{c2}$. However, as the correction in $\chi_s$ does not affect the order of $\chi_n - \chi_s$, the enhancement of $B_{c2}$ is not affected in a qualitative way given $B_{c2} = B_p \sqrt{\chi_0 / (\chi_n - \chi_s)}$ as we discussed in the main text.

%In the absence of SOC, we have $\braket{a_{ii}(E_F)}=1, \braket{a_{ij\neq i}(E_F)}=0$ and
%\begin{equation}
% \overline{\chi_s^{ii}}/\chi_n^{ii}=\begin{cases}
% 1-\pi k_BT\sum_n \frac{\Delta^2}{(\omega_n^2+\Delta^2)^{3/2}},\ i=z\\
%  1-\pi k_BT\sum_n \frac{\Delta^2}{(\omega_n^2+\Delta^2)^{3/2}}\frac{1-\frac{1}{\tau_{so}\sqrt{\omega_n^2+\Delta^2}+2}}{1+\frac{1}{\tau_{so}\sqrt{\omega_n^2+\Delta^2}}},\ i=x,y\label{spin_sus_spinorbit}
% \end{cases}
%\end{equation}

In conclusion, we find that $B_{c2}$ in the SOPC superconductor is robust against scalar potential fluctuations and spin-orbit scattering may further enhance $B_{c2}$ by inducing a residue $\chi_{s}$.

\section{Possibility of inter-orbital pairings}

In the main text, we assumed \textit{intra-orbital} pairing which is expected to be favored when the intra-orbital attractive interaction dominates. However, given that the monolayer WTe$_2$ becomes superconducting near the topological band crossing points where different orbitals are strongly mixed by SOPC, instability toward \textit{inter-orbital} pairings is also possible under inter-orbital attractive interactions and worth to be explored. In the following, we study the following properties of possible inter-orbital pairings: (i) symmetry classification, (ii) pairing instability, (iii) topological nature, and (iv) enhancement of $B_{c2}$. In particular, we discuss the important role of SOPC in these special properties of inter-orbital pairing.

\subsection{Symmetry classification}

In the Nambu basis $(c_{\bm{k},\uparrow},c_{\bm{k},\downarrow},c^{\dagger}_{-\bm{k},\downarrow},-c^{\dagger}_{-\bm{k},\uparrow})^{T}$ with $c_{\bm{k},\sigma}=(c_{p,\bm{k},\sigma},c_{d,\bm{k},\sigma})^{T}$, the pairing matrix transforms as \cite{FuliangS,VenderbosS}
\begin{eqnarray}
T: \hat{\Delta}(\bm{k}) & \mapsto& \sigma_y\hat{\Delta}^{*}(-\bm{k})\sigma_y;\\
g: \hat{\Delta}(g\bm{k}) & \mapsto& U(g)\hat{\Delta}(\bm{k})U^{-1}(g)
\end{eqnarray}
where $T=i\sigma_yK$ is the time-reversal operation,  $g$ is a symmetry operation in the $C_{2h}$ point group of monolayer WTe$_2$. In our convention, the mirror operation $\sigma_h$ defined in the usual character table of $C_{2h}$ is the mirror reflection about the $xz$-plane $M_y: (x,y,z) \mapsto (x,-y,z)$. By imposing time-reversal-symmetry and fermi statistics, all possible intra-unit-cell (\textit{i.e.}, $\bm{k}$-independent) pairing matrices are listed in Table \ref{TableS3} below and classified according to the irreducible representations (IRs) of $C_{2h}$. 

\begin{table}[h]
	\caption{Classifications of all time-reversal-invariant intra-unit-cell pairings according to the irreducible representations (IRs) of $C_{2h}$ point group for monolayer WTe$_2$. The pairings are written in matrix form under the Nambu basis $(c_{\bm{k},\uparrow},c_{\bm{k},\downarrow},c^{\dagger}_{-\bm{k},\downarrow},-c^{\dagger}_{-\bm{k},\uparrow})^{T}$ with $c_{\bm{k},\sigma}=(c_{p,\bm{k},\sigma},c_{d,\bm{k},\sigma})^{T}$.}
	\begin{tabular}{cccc}
		\hline\hline
		IRs &\hspace{1 mm} $A_g$  &\hspace{1 mm} $A_u$ &\hspace{1 mm} $B_u$\\\hline	
		$P$ &\hspace{1 mm} $+$ &\hspace{1 mm} $-$ &\hspace{1 mm} $-$\\
		$M_y$ &\hspace{1 mm} $+$ &\hspace{1 mm} $-$ &\hspace{1 mm} $+$\\\hline
		Singlet &\hspace{1 mm} $\eta_1 s_0$, $\eta_1 s_z$ &\hspace{1 mm} None &\hspace{1 mm} $\eta_1 s_x$\\
		Triplet  &\hspace{1 mm}   None            &\hspace{1 mm} $\eta_1 s_y\sigma_x$, $\eta_1 s_y\sigma_z$ &\hspace{1 mm} $\eta_1 s_y\sigma_y$\\\hline
		\hline
	\end{tabular}
	\label{TableS3}
\end{table}
The trivial $A_g$ phase describes the intra-orbital spin-singlet pairing we considered in the main text. The nontrivial $A_u$ phase includes two inter-orbital triplet pairings $\hat{\Delta}_{A_u,1} = \eta_1 s_y\sigma_x$, $ \hat{\Delta}_{A_u,2} = \eta_1 s_y\sigma_z$, while the other nontrivial $B_u$ phase includes one inter-orbital spin-singlet pairing $\hat{\Delta}_{B_u, 1} =\eta_1 s_x$ and one inter-orbital spin-triplet pairing $\hat{\Delta}_{B_u, 2} = \eta_1 s_y\sigma_y$, respectively.

\subsection{Pairing instability}

\begin{figure}
	\centering
	\includegraphics[width=0.9\linewidth]{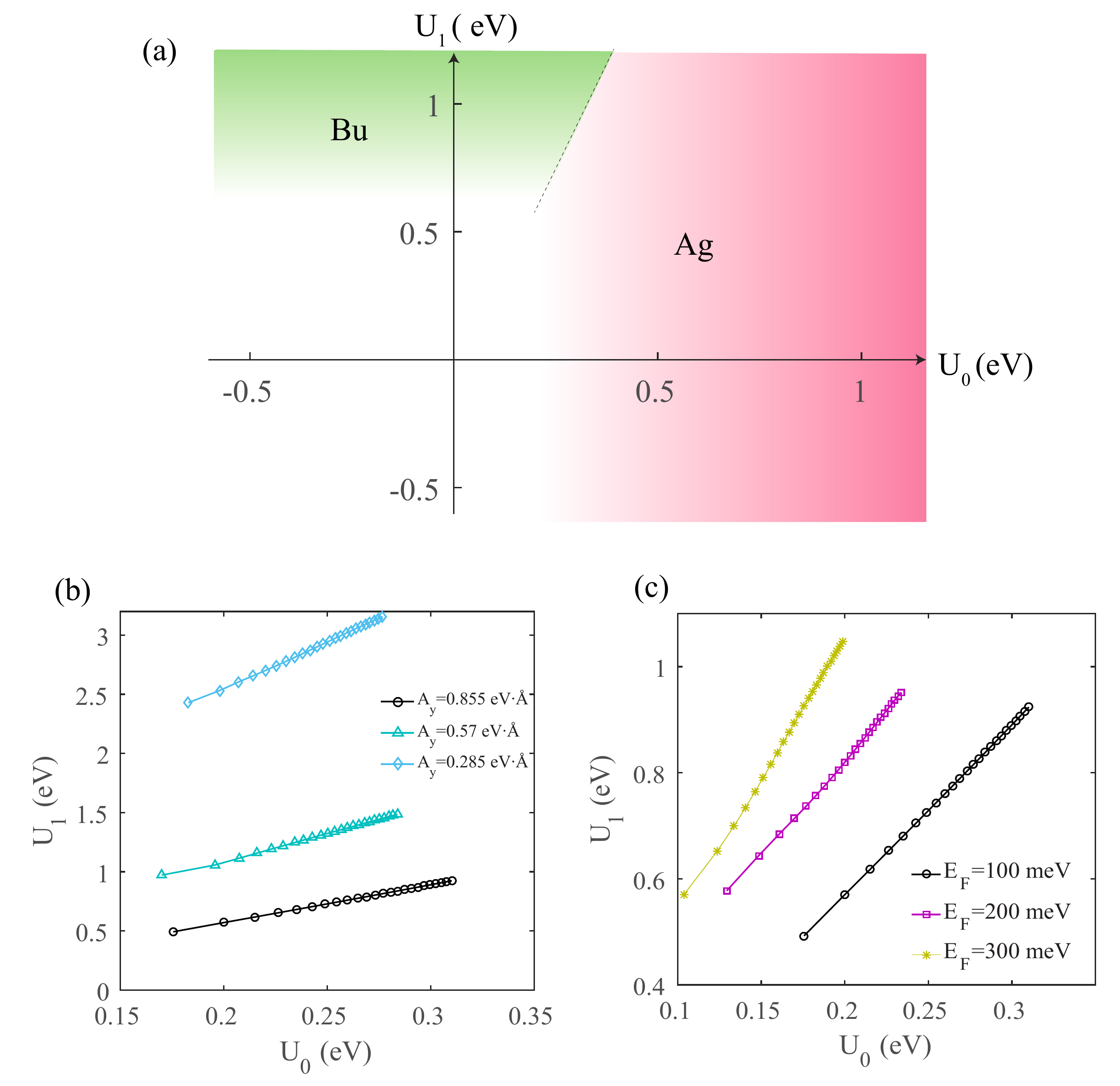}
	\caption{(a) Superconducting phase diagram with intra-orbital $\hat{\Delta}_0$ ($A_g$-phase) and inter-orbital pairing $\hat{\Delta}_1$ ($B_u$-phase). $U_i>0$ denotes attractive interaction ($i=0/1$ stands for intra/inter-orbital). (b)-(c) Phase boundaries between $B_u$/$A_g$ pairings under different SOPC strengths $A_y$ and chemical potential $E_F$. Black dotted lines in (b)-(c) are identical to the phase boundary in (a). By reducing $A_y$ (b) or increasing $E_F$ (c), the SOPC effect is reduced and a larger $U_1$ is generally required for $B_u$ phase to be favored.}
	\label{fig:figs4}
\end{figure}

To study the pairing instability under attractive interactions, we start from the general interacting Hamiltonian:
\begin{equation}\label{Ham7}
\hat{H}_{int}=\frac{1}{2}\sum_{\bm{p},\bm{p'}}V_{\alpha\beta\gamma\delta}^{ijkl} \psi_{i,\alpha,-\bm{p}}^{\dagger}\psi^{\dagger}_{j,\beta,\bm{p}}\psi_{k,\gamma,\bm{p'}}\psi_{l,\delta,-\bm{p'}}.
\end{equation}
%We set the intra-orbital singlet interaction is $V^{iiii}_{\alpha\beta\beta\alpha}=V^{iiii}_{\alpha\beta\alpha\beta}=-U_0$ ($\alpha\neq \beta$), the inter-orbital  interaction is $V^{ijji}_{\alpha\alpha\alpha\alpha}=-V^{ijij}_{\alpha\alpha\alpha\alpha}=-V_0, V^{ijji}_{\alpha\beta\beta\alpha}=-V^{ijij}_{\alpha\beta\alpha\beta}=-V_0'$ ($\alpha\neq \beta$).
where $i,j,k,l$ and $\alpha,\beta,\gamma,\delta$ are the orbital and spin indices respectively. With $\hat{H}_{int}$ respecting all point group $g \in C_{2h}$, time-reversal and SU(2) symmetries, $V_{\alpha\beta\gamma\delta}^{ijkl}$ can be decomposed into different channels as
\begin{equation}
V_{\alpha\beta\gamma\delta}^{ijkl}=-\sum_{\Gamma,m}V_{\Gamma,m}(\hat{\Delta}_{\Gamma,m}i\sigma_y)_{ij,\alpha\beta}(\hat{\Delta}^{\dagger}_{\Gamma,m}i\sigma_y)_{kl,\gamma\delta}.
\end{equation}
%By comparing with the interactions in Eq.~\ref{Ham7}, it can be found the intra-orbital interaction $V_{A_g,1}=V_{A_g,2}=U_0$, the inter-orbital interaction  $V_{A_u,1}=V_{B_u,2}=V'_0, V_{A_u,2}=V_{B_u,1}=V_0$. 
Here, $\Gamma$ labels different irreducible representations(IRs), $m$ labels the possible components in each $\Gamma$, and the forms of  $\hat{\Delta}_{\Gamma,m}$ for a given $(\Gamma, m)$ corresponds to one particular component in Table \ref{TableS3}. Note that the minus sign in front of the summation captures the attractive nature of the interaction, thus $V_{\Gamma,m}>0(<0)$ denotes attractive(repulsive) interaction in the given channel labelled by $(\Gamma, m)$. In each pairing phase belonging to a representation $\Gamma$, the critical temperature is given by
\begin{equation}
\text{det}\begin{bmatrix}
\begin{pmatrix}
V_{\Gamma,1}\chi_{\Gamma, 11}(T_c)&V_{\Gamma,1}\chi_{\Gamma, 12}(T_c)\\
V_{\Gamma,2}\chi_{\Gamma, 21}(T_c)&V_{\Gamma,2}\chi_{\Gamma, 22}(T_c)
\end{pmatrix}-I
\end{bmatrix}=0,\label{eq:pairingeq}
\end{equation}
where $\chi_{\Gamma ,mm'}$ denotes the pairing susceptibility:
\begin{eqnarray}
\chi_{\Gamma ,mm'}&&=-\frac{1}{\beta}\sum_{n,\bm{p}}\text{Tr}(G_e(\bm{p},i\omega_n)\Delta_{\Gamma,m}G_h(\bm{p},i\omega_n)\Delta^{\dagger}_{\Gamma,m'})\nonumber\\
&&=\int \frac{d^2\bm{p}}{(2\pi)^2} \sum_{a,b}O^{\Gamma m}_{a,b}(\bm{p})O^{\Gamma m'}_{a,b}(\bm{p})\frac{1-f(E_a(\bm{p})-f(E_b(-\bm{p}))}{E_a(\bm{p})+E_b(-\bm{p})}.
\end{eqnarray}
Here, the single particle electron Green's function $G_e(\bm{p},i\omega_n)=(i\omega_n-H_0(\bm{p}))^{-1}$ and hole Green's function $G_h(\bm{p},i\omega_n)=(i\omega_n+H_0(\bm{p}))^{-1}$, the overlap function $O^{\Gamma m}_{a,b}(\bm{p})=\braket{u_{a,\bm{p}}|\Delta_{\Gamma,m}|\nu_{b,\bm{p}}}$ with $\ket{u_{a,\bm{p}}}$,$\ket{\nu_{b,\bm{p}}}$ being eigenvectors of $H_0(\bm{p})$ satisfying $H_0(\bm{p})\ket{u_{a,\bm{p}}}=E_a(\bm{p})\ket{u_{a,\bm{p}}}, H_0(\bm{p})\ket{\nu_{b,\bm{p}}}=E_b(\bm{p})\ket{\nu_{b,\bm{p}}}$, $a,b$ are the band indices.

To further simplify our analysis, we note that the experimentally observed $B_{c2}$ is only $1-3$ times higher than the Pauli limit for $T \rightarrow 0$ \cite{FatemiS, SajadiS}, which is not compatible with triplet pairing phases: for $A_u$ phase, the combination of two triplet pairings $\hat{\Delta}_{A_u,1} = \eta_1 s_y\sigma_x$ and $ \hat{\Delta}_{A_u,2} = \eta_1 s_y\sigma_z$ are characterized by a triplet $\bm{d}$-vector of the general form $\bm{d} = (d_x, 0, d_z)$, which is parallel to the $xz$-plane and generates equal-spin Cooper pairs with spins in the $y$-direction \cite{Zhou}. This would lead to large superconducting spin susceptibility \cite{Sigrist_susceptibility} and a divergent $B_{c2}$ for fields along the $y$-direction as $T \rightarrow 0$. This motivates us to first rule out the $A_u$ phase. 

On the other hand, the $B_u$ phase also has a triplet component $\hat{\Delta}_{B_u,2} = \eta_1 s_y\sigma_y$ with the triplet $\bm{d}$-vector: $\bm{d} = (0, d_y, 0)$, which generates equal-spin Cooper pairs with spins parallel to the $xz$-plane and leads to divergent $B_{c2}$ for fields along the $x$-direction as $T \rightarrow 0$. As the singlet $\hat{\Delta}_{B_u,1} = \eta_1 s_x$ and triplet $\hat{\Delta}_{B_u,2} = \eta_1 s_y\sigma_y$ components in $B_u$ phase can mix in general, the discrepancy between $\hat{\Delta}_{B_u,2}$ and the experimental observation further motivates us to consider the channel dominated by $\hat{\Delta}_{B_u,1} = \eta_1 s_x$. In fact, numerically we find that $\chi_{\Gamma, 11} \approx \chi_{\Gamma, 22}$ and the singlet-triplet mixing $\chi_{\Gamma, 12} = \chi^{*}_{\Gamma, 21}$ between $\hat{\Delta}_{B_u,1}$ and $\hat{\Delta}_{B_u,2}$ is negligibly small. Therefore, there does exist a singlet-dominant phase in $B_u$ if $V_{B_u,1}$ dominates over $V_{B_u,2}$. We note that the condition $V_{B_u,1} \gg V_{B_u,2}$ can indeed be met under realistic considerations: by projecting the general interaction $\hat{H}_{int}$ to the $\hat{\Delta}_{B_u,1}$ and $\hat{\Delta}_{B_u,2}$ channels, it can be shown explicitly that: $V_{B_u,1} =- (I_{pd} + J_{pd})$ and $V_{B_u,2} =- (I_{pd} - J_{pd})$, where $I_{pd}$ and $J_{pd}$ stand for the inter-orbital direct coupling and exchange coupling terms given by:
\begin{eqnarray}
I_{pd} &=& \int d\bm{r} d\bm{r}' |\phi_{p} (\bm{r})|^{2} V(|\bm{r}-\bm{r}'|)  |\phi_{d} (\bm{r}')|^{2}, \\\nonumber
J_{pd} &=& \int d\bm{r} d\bm{r}' \phi^{*}_{p} (\bm{r}) \phi^{*}_{d} (\bm{r}')V(|\bm{r}-\bm{r}'|) \phi_{p} (\bm{r}') \phi_{d} (\bm{r}),
\end{eqnarray}
where $\phi_{l =p,d} (\bm{r})$ describes the Wannier orbital with $p,d$ characters localized within the unit cell, and $ V(|\bm{r}-\bm{r}'|) < 0$ describes the microscopic attractive interaction leading to pairing instability. To drive pairing instability toward inter-orbital pairing, the spatial overlap between Wannier $p,d$-orbitals is required to be strong. Thus, one expects $I_{pd}\sim J_{pd} <0$, and $V_{B_{u,1}} =- (I_{pd} + J_{pd}) \gg V_{B_{u,2}} =- (I_{pd} - J_{pd})$, which simply reflects the fact that singlet states generally acquire a larger attractive interaction strength due to its symmetrical orbital part of the two-body wave function \cite{Tinkham}. 

Based on the observations above, we focus on the inter-orbital singlet $\hat{\Delta}_{B_u,1}$ pairing and compare its pairing instability with the intra-orbital singlet $A_g$ phase considered in the main text. For simplicity of the following discussions, we relabel the intra-orbital interaction as $V_{A_g,1}=V_{A_g,2}=U_0$, the inter-orbital interaction as $V_{B_u,1}=U_1$, and we refer to the inter-orbital singlet $\hat{\Delta}_{B_u,1}$ pairing phase simply as the $B_u$ phase.

The superconducting phase diagram with $B_u$ and $A_g$ pairing phases is shown in Fig.~\ref{fig:figs4}, where the more favorable phase at a given point ($U_0$, $U_1$) is determined by the phase with highest $T_c$. Here, $U_{i=0,1}>0$ ($U_{i=0,1}<0$) denotes the interaction being attractive (repulsive). The chemical potential is set to be close to the topological band crossing points as in Fig.~2 of the main text. When inter-orbital interaction is repulsive ($U_1<0$), an intra-orbital attraction $U_0>0$ leads to instability toward intra-orbital $A_g$ phase. In contrast, when intra-orbital interaction is repulsive $(U_0<0)$, an inter-orbital attraction $U_1>0$ leads to instability toward the inter-orbital $B_u$ phase. In the regime where $U_0,U_1>0$, the $A_g(B_u)$-phase is more energetically favored when $U_0 (U_1)$ dominates. As the two pairing phases belong to different irreducible representations, these two pairings do not mix, and a phase transition happens at the well-defined phase boundary indicated by the dashed line in Fig.~\ref{fig:figs4}a. 

Notably, the inter-orbital $B_u$-phase is sensitive to the SOPC as the effective pairing strength is controlled by the mixing between $p,d$-orbitals. By fixing the chemical potential near the band crossing point and reducing the SOPC strength $A_y$ gradually, the phase boundary between $B_u$-phase and $A_g$-phase gets shifted upward (Fig.~\ref{fig:figs4}b). This indicates that a stronger interaction $U_1$ is needed for the inter-orbital $B_u$-pairing phase to be favored. Moreover, as we discussed in the main text, the SOPC effect is only important near the topological band crossing points. Thus, by tuning the chemical potential away from the band crossing points, the SOPC effect is reduced. In this case, the phase boundary also gets shifted upward with the regime favoring the $B_u$-pairing phase being reduced (Fig.~\ref{fig:figs4}c). These results clearly show that the SOPC helps to stabilize the $B_u$-phase under inter-orbital attractive interactions.

\subsection{Topological nature}

It is interesting to note that the $B_u$ phase is an odd-parity pairing phase (Table \ref{TableS3}), similar to the odd-parity pairing studied in Cu-doped Bi$_2$Se$_3$ \cite{FuliangS}. As we pointed out in the main text, this odd-parity pairing results in a DIII class topological superconductor when the Fermi surface encloses odd number of time-reversal-invariant (TRIM) points \cite{FuliangS}. Indeed, given a nonzero mean-field order parameter $\Delta_1$ for the $B_u$ phase, the explicit form of $B_u$ pairing can be written as: $\hat{\Delta}_1 = \Delta_1 ( c^{\dagger}_{\bm{k}, p, \uparrow} c^{\dagger}_{-\bm{k}, d, \downarrow} - c^{\dagger}_{\bm{k}, p, \downarrow} c^{\dagger}_{-\bm{k}, d, \uparrow} + h.c.)$. We explicitly reveal the nontrivial topological nature of the $B_u$ pairing below by showing that projecting $\hat{\Delta}_1$ to the MCPB basis results in an effective $p\pm ip$ pairing. For simplicity, we drop $A_z k_y$ terms in the SOPC given $A_y  , A_x \gg A_z$ as shown in Table \ref{table2}.

In the basis of $\ket{\bm{k}, p,\uparrow}, \ket{\bm{k}, p,\downarrow}, \ket{\bm{k}, d,\uparrow}, \ket{\bm{k}, d,\downarrow}$, the psedospin basis is given by:
\begin{eqnarray}
\ket{\bm{k},\alpha} = \frac{1}{2 N_{\bm{k}} }
\begin{pmatrix}
P(\bm{k}) (e^{i \frac{\alpha_{\bm{k}} }{2} } + e^{- i \frac{\alpha_{\bm{k}} }{2} }) \\
e^{i \phi_{\bm{k}}} P(\bm{k})  (e^{i \frac{\alpha_{\bm{k}} }{2} } - e^{- i \frac{\alpha_{\bm{k}} }{2} })\\
D(\bm{k}) e^{i \frac{\alpha_{\bm{k}} }{2} } - D^{*} (\bm{k}) e^{-i \frac{\alpha_{\bm{k}} }{2} } \\
e^{i \phi_{\bm{k}}} (D(\bm{k}) e^{i \frac{\alpha_{\bm{k}} }{2} } + D^{*} (\bm{k}) e^{-i \frac{\alpha_{\bm{k}} }{2} } )
\end{pmatrix}
, && \ket{\bm{k},\beta} = \frac{1}{2 N_{\bm{k}} }
\begin{pmatrix}
e^{-i \phi_{\bm{k}}} P(\bm{k})  (e^{i \frac{\alpha_{\bm{k}} }{2} } - e^{- i \frac{\alpha_{\bm{k}} }{2} })\\
P(\bm{k}) (e^{i \frac{\alpha_{\bm{k}} }{2} } + e^{- i \frac{\alpha_{\bm{k}} }{2} })\\
e^{-i \phi_{\bm{k}}} (D(\bm{k}) e^{i \frac{\alpha_{\bm{k}} }{2} } + D^{*} (\bm{k}) e^{-i \frac{\alpha_{\bm{k}} }{2} } )\\
D(\bm{k}) e^{i \frac{\alpha_{\bm{k}} }{2} } - D^{*} (\bm{k}) e^{-i \frac{\alpha_{\bm{k}} }{2} } 
\end{pmatrix},
\end{eqnarray}
where $e^{i \phi_{\bm{k}}} = (A_y k_y + i A_x k_x)/Ak$, $P(\bm{k}) = E(\bm{k}) + \mathcal{M}(\bm{k})$ and $D(\bm{k}) = i v k_x + A k$ characterize the weights of the $p$ and $d$-orbitals in the psedospin basis. By defining $f_{p,+}(\bm{k}) = P(\bm{k}) (e^{i \frac{\alpha_{\bm{k}} }{2} } + e^{- i \frac{\alpha_{\bm{k}} }{2} })$, $ f_{p,-}(\bm{k}) = -i P(\bm{k})  (e^{i \frac{\alpha_{\bm{k}} }{2} } - e^{- i \frac{\alpha_{\bm{k}} }{2} })$, $f_{d,+}(\bm{k}) = (D(\bm{k}) e^{i \frac{\alpha_{\bm{k}} }{2} } + D^{*} (\bm{k}) e^{-i \frac{\alpha_{\bm{k}} }{2} } )$, $f_{d,-}(\bm{k}) = -i( D(\bm{k}) e^{i \frac{\alpha_{\bm{k}} }{2} } - D^{*} (\bm{k}) e^{-i \frac{\alpha_{\bm{k}} }{2} } )$ [note: $f_{p,\pm}(\bm{k}), f_{\pm}(d,\bm{k})$ are all real functions of $\bm{k}$, with $f_{p,\pm}(- \bm{k}) = \pm f_{p,\pm}(\bm{k}), f_{d,\pm}(- \bm{k}) = \pm f_{d,\pm}( \bm{k})$], we project $c^{\dagger}_{\bm{k}, l, \sigma} =  \braket{\bm{k}, \alpha|\bm{k},  l, \sigma } c^{\dagger}_{\bm{k},\alpha} +  \braket{\bm{k}, \beta|\bm{k},  l, \sigma} c^{\dagger}_{\bm{k},\beta} $ with $l=p,d, \sigma = \uparrow, \downarrow$, and $\hat{\Delta}_1$ is reduced to:
\begin{equation}\label{eq:EffectiveP}
\hat{\Delta}_{1, eff} (\bm{k}) = \Delta_1\frac{f_{p,-}(\bm{k}) f_{d,-}(\bm{k}) + f_{p,+}(\bm{k}) f_{d,+}(\bm{k})}{4 N^2_{\bm{k}} Ak} \left[\left(A_yk_y-iA_xk_x\right)c_{\bm{k},\alpha}^\dag c_{-\bm{k},\alpha}^\dag-\left(A_yk_y+iA_xk_x\right)c_{\bm{k},\beta}^\dag c_{-\bm{k},\beta}^\dag\right].
\end{equation}

Clearly, $\hat{\Delta}_{1, eff} (\bm{k})$ reveals that the combined effect of $\hat{\Delta}_1$ and SOPC leads to an effective $p \pm ip$ pairing. Notably, in the absence of SOPC ($A_y = A_x=0$) or $\hat{\Delta}_1$ ($\Delta_1 = 0$), the effective $p$-wave gap function $\hat{\Delta}_{1, eff} (\bm{k})$ vanishes and the bulk spectrum of Bogoliubov quasi-particles remains gapless. Thus, when the Fermi surface encloses odd number of TRIM-points, the effective $p\pm ip$ pairing leads to a time-reversal-invariant topological superconductor.

\begin{figure}
	\centering
	\includegraphics[width=0.9\linewidth]{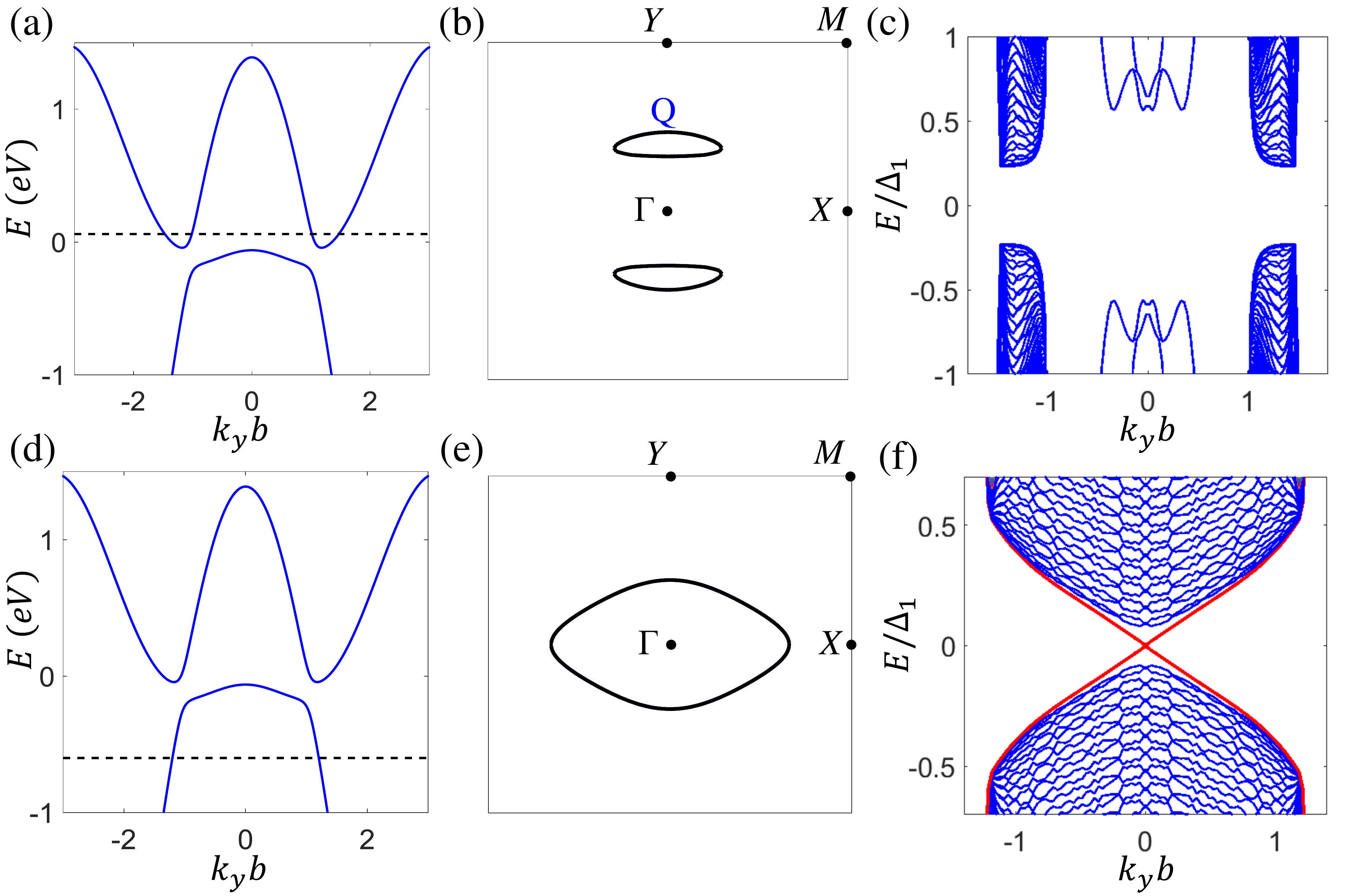}
	\caption{Spectrum of a strip of superconducting WTe$_2$ in $B_u$ pairing phase with pairing amplitude $\Delta_1$. (a) Under realistic conditions, WTe$_2$ becomes superconducting when conduction band states near the Q-points are filled, where the Fermi surface is formed by two disconnected Q-pockets enclosing none of the TRIM-points (b). In this case, no helical Majorana modes can form on the edge (c). By artificially tuning the chemical potential to the hole band (d) such that the $\Gamma$-point is enclosed by the Fermi surface (e), the system becomes a DIII class topological superconductor with helical Majorana modes on the edge (f). Details of the tight-binding model used in (a)-(f) are presented in subsection E of this Supplementary Material.}
	\label{fig:figs6}
\end{figure}

Unfortunately, as we pointed out in the main text, superconductivity in monolayer WTe$_2$ sets in when the conduction bands near Q-points are filled (Fig.~\ref{fig:figs6}a), where the Fermi surface consists of two disconnected Fermi pockets enclosing none of the four TRIM points $\mathrm{\Gamma}$,X,Y,Z (Fig.~\ref{fig:figs6}b). As a result, the system remains topologically trivial. To demonstrate this explicitly, we use a tight-binding model to calculate the energy spectrum of a finite WTe$_2$ strip under $\hat{\Delta}_1$. Clearly, no helical Majorana modes can form on the edge as shown in Fig.~\ref{fig:figs6}c. To reveal the nontrivial nature of $\hat{\Delta}_1$, we artificially tune the chemical potential to the hole bands (Fig.~\ref{fig:figs6}d) such that the $\Gamma$-point is enclosed by the Fermi surface (Fig.~\ref{fig:figs6}e). In this case, helical Majorana states emerge on the edge (Fig.~\ref{fig:figs6}f), which clearly shows that the system becomes a DIII class topological superconductor. Details of the tight-binding model used to obtain the edge state spectrum in Fig.~\ref{fig:figs6} are presented in subsection E of this Supplementary Material.

As an explanatory note, we point out that the mechanism behind the generation of effective $p$-wave pairing from a singlet-pairing $\hat{\Delta}_1$ in the orbital-basis is similar to the effective $p$-wave pairing created by $s$-wave pairing and strong noncentrosymmetric spin-orbit couplings(SOCs) \cite{Sigrist_susceptibility, Zhou, Alicea}. From the symmetry point of view, such phenomena arises from the breaking of both inversion and $SU(2)$ spin-rotation symmetries. In the case of noncentrosymmetric superconductors, the noncentrosymmetric SOC in the normal state breaks both inversion and $SU(2)$, while in the case of the $B_u$ phase in superconducting WTe$_2$, the inversion-breaking due to $\hat{\Delta}_1$ and $SU(2)$-breaking from SOPC work together to produce the effective $p$-wave pairing. 

Moreover, the odd-parity nature of $\hat{\Delta}_1$ forbids any pseudospin-singlet pairing in the effective pairing Hamiltonian: given a general effective pairing $\hat{\Delta}_{eff}(\bm{k}) = \psi (\bm{k}) \rho_0 + \bm{d} (\bm{k}) \cdot \bm{\rho}$ under pseudospin basis, partiy transforms $\hat{\Delta}_{eff}(\bm{k})$ as: $\hat{\Delta}_{eff}(\bm{k}) \mapsto \hat{\Delta}_{eff}(-\bm{k})$, while fermi statistics requires $\psi (\bm{k}) = \psi (-\bm{k}), \bm{d} (\bm{k}) = - \bm{d} (-\bm{k})$. As the odd-parity condition imposes $\hat{\Delta}_{eff}(\bm{k}) = - \hat{\Delta}_{eff}(-\bm{k})$, the pseudospin singlet component is forced to vanish: $\psi (\bm{k}) = 0$. As we discuss next, the pseudospin triplet component in $\hat{\Delta}_{1, eff}(\bm{k})$ under the odd-parity $B_u$ phase has important consequences on the spin magnetic properties of superconducting WTe$_2$.

\subsection{Enhancement of $B_{c2}$}

\begin{figure}
	\centering
	\includegraphics[width=0.9\linewidth]{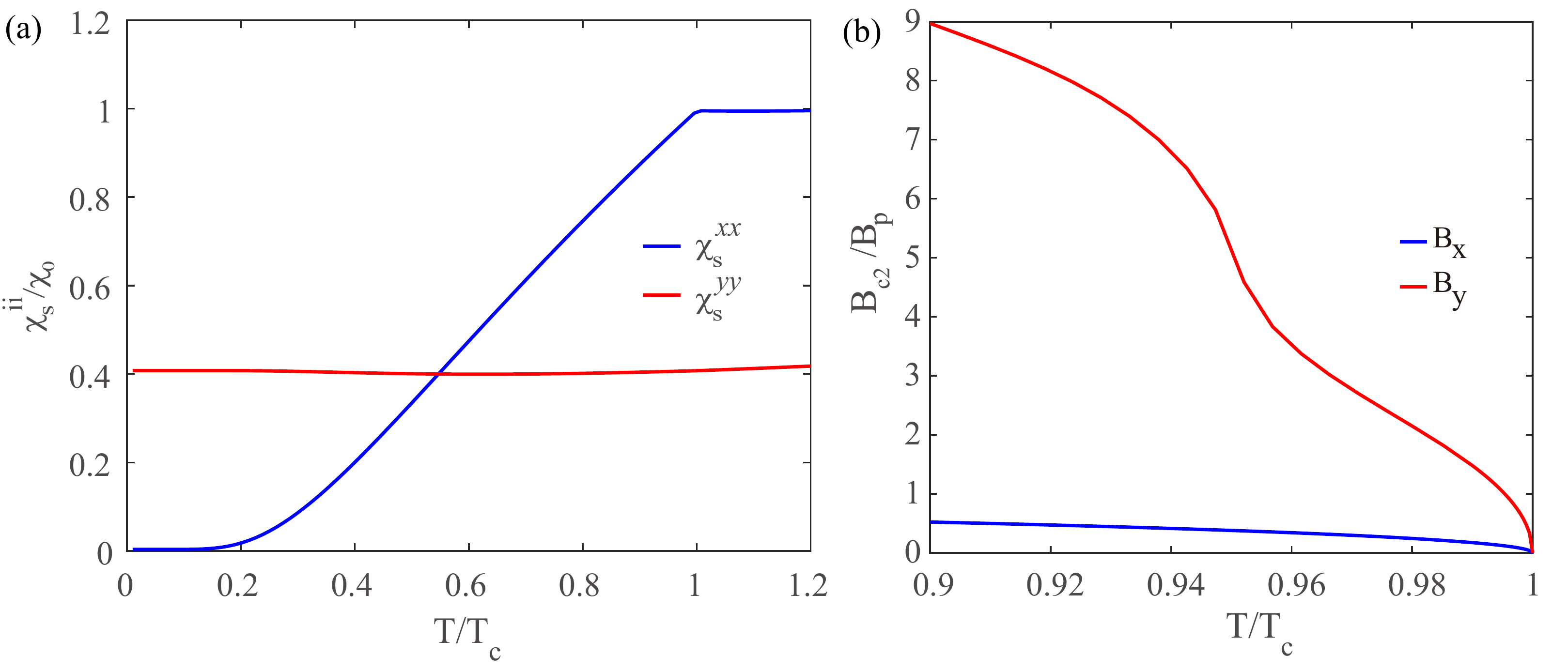}
	\caption{Spin susceptibility and upper critical  field for the $B_{u}$ pairing. (a) Superconducting spin susceptibility $\chi_s^{ii}$ $(i=x,y)$ as a function of temperature $T$. (b) The upper critical field $B$ along $x,y$- direction as a function of temperature $T$ obtained from solving linearized gap equations numerically. Except for the pairing form, other parameters are the same with the Fig.~3a of main text.   }
	\label{fig:figs5}
\end{figure}

Finally, we show how $\hat{\Delta}_1$ under $B_{u}$ pairing phase affects the in-plane $B_{c2}$. As we discussed in the main text, when states near $Q$-points are filled, there is a large anisotropy in the SOPC: $A_yk_y\gg A_xk_x, A_zk_x\sim0$. In other words, the $k_y$-component in the effective $p\pm ip$ pairing dominates near $Q$-points (Eq.\ref{eq:EffectiveP}). This allows us to approximately set $\theta_{\bm{k}} = \pi/2, \sin \phi_{\bm{k}}=0$ and the components in the $\bm{d}$-vector under pseudospin basis are given by:
\begin{equation}\label{eq:PairCorrelationPseudo}
d_x(\bm{k})=-\Delta_1 (E(\bm{k})+\mathcal{M}(\bm{k}))A_yk_y/N_{\bm{k}}^2, \hspace{1 mm} d_y(\bm{k})=d_z(\bm{k})=0.
\end{equation}

Notably, the spin magnetic property of the superconducting state is determined by the spin structure of pairing electrons in the real spin basis. To see how $\hat{\Delta}_1$ affects the spin properties of Cooper pairs formed by electrons near the Fermi surface, we need to study how the real-spin triplet $\bm{\tilde{d}}$-vector is related to the pseudospin triplet $\bm{d}$-vector in Eq.\ref{eq:PairCorrelationPseudo}. Recall that any component $\tilde{d}_i  (i=x,y,z)$ of a real-spin triplet $\bm{\tilde{d}}$-vector is given by: $\tilde{d}_i(\bm{k}) = \textrm{Tr}[ \sigma_i \hat{\Delta}_t(\bm{k}) ]/2$, where $\hat{\Delta}_t(\bm{k}) = \bm{\tilde{d}}(\bm{k})\cdot \bm{\sigma}$ is the usual reap-spin triplet pairing characterized by a nonzero $\bm{\tilde{d}}$ and $\sigma_i$ are the Pauli matrices for real spins. Therefore, in the pseudospin basis representation, the components of real-spin $\tilde{\bm{d}}$ is given by 
\begin{equation}
\tilde{d}_i(\bm{k}) = \frac{1}{2}\textrm{Tr} [\tilde{\sigma}_{i}(\bm{k}) \hat{\Delta}_{1, eff} (\bm{k})] \approx a_{ix}(\bm{k}) d_x (\bm{k}),
\end{equation}
where $\tilde{\sigma}_{i} (\bm{k})$ are the real-spin Pauli matrices under pseudospin basis presented in Eq.\ref{eq:PauliPseudospin}, and $a_{ix}(\bm{k})$ are the coefficients associated with $\sigma_{i}$ and $\rho_x$. Notably, with the approximation $\theta_{\bm{k}} = \pi/2, \sin \phi_{\bm{k}}=0$, we have $\tilde{\sigma}_x = \rho_x, \tilde{\sigma}_y(\bm{k}) = |W_{\bm{k}}|\rho_y, \tilde{\sigma}_z(\bm{k}) = |W_{\bm{k}}|\rho_z$, where $W_{\bm{k}}$ is defined in Eq.\ref{eq:PauliPseudospin}. As such, we have $a_{xx} = 1$, $a_{yx} = a_{zx} = 0$, and the real-spin triplet $\bm{\tilde{d}}$-vector in this particular case is indeed almost identical to the pseudospin triplet $\bm{d}$-vector: $\bm{\tilde{d}} \approx \bm{d} = (d_x(\bm{k}), 0, 0)$ which has a nonzero $\tilde{d}_x(\bm{k})$-component. This nonzero $\tilde{d}_x(\bm{k})$ generates equal-spin Cooper pairs with spins parallel to the $yz$-plane, similar to the case of Ising superconductors where the nonzero $d_z(\bm{k})$ due to Ising SOC generates equal-spin Cooper pairs with spins parallel to the $xy$-plane \cite{Zhou}. Therefore, under $\hat{\Delta}_{1}$ pairing there exists a large $\chi^{yy}_{s}$ in the superconducting state and the in-plane $B_{c2}$ along $y$-direction is expected to be enhanced much more dramatically and diverge in the $T \rightarrow 0$ limit.

To verify our analysis above based on the pseudospin basis, we explicitly demonstrate the effect of $\hat{\Delta}_{1}$ on the spin magnetic properties and $B_{c2}$ of the superconducting WTe$_2$. To be specific, based on the full $H_{BdG}(\bm{k})$ in Eq.\ref{eq:FullBdG} with the pairing matrix replaced by $\hat{\Delta}_{1} = \Delta_1 \eta_1 s_x$, we calculate the superconducting spin susceptibility numerically using the Kubo formula 
 \begin{align}
\chi_s^{ii}=-\frac{1}{2}u_B^2\lim_{\bm{q}\rightarrow 0}\sum_{\bm{k},m= n}\sum_{a,b} \frac{f(E_m(\bm{k}))-f(E_n(\bm{k}+\bm{q}))}{E_m(\bm{k})-E_n(\bm{k}+\bm{q})}\braket{n,\bm{k},b|\sigma^{i}|m,\bm{k},a}\braket{m,\bm{k},a|\sigma^{i}|n,\bm{k},b},
\end{align}
where $ f(E)$ is the Fermi distribution function,  eigenenergies $E_{n}(\bm{k})$ and eigenstates $\ket{n,\bm{k},a}$ are calculated from the full four-band BdG Hamiltonian $H_{BdG}(\bm{k})$ at zero field, $a=1,2$ labels the two degenerate states. The superconducting spin susceptibility $\chi^{xx}_s$ and $\chi^{xx}_s$ under $\hat{\Delta}_1$ pairing as a function of temperature $T$ are shown in Fig.~\ref{fig:figs5}. Notably, the superconducting spin susceptibility $\chi_s^{xx}$ along the $x$-direction under $\hat{\Delta}_1$ is similar to the case with intra-orbital $A_g$ pairing in Fig.3 of the main text (blue line in Fig.~\ref{fig:figs5}a). This is because $\tilde{d}_x(\bm{k})$ generates no equal-spin Cooper pairs with spins pointing to the $x$-direction. Consistently, $B_{c2}$ under $B_x$ is also similar to the case with intra-orbital pairing (blue line in Fig.~\ref{fig:figs5}b). In sharp contrast, $\chi_s^{yy}$ along the $y$-direction is non-vanishing in the $T \rightarrow 0$ limit, and remains nearly the same as its normal state value $\chi_n^{yy}$ (red line in Fig.~\ref{fig:figs5}a), indicating the superconductivity is insensitive to the in-plane field along the $y$-direction. Consequently, $B_{c2}$ under $B_y$ gets dramatically enhanced in the $\hat{\Delta}_1$ phase belonging to the $B_u$ representation, which easily exceeds the Pauli limit by nearly ten times even in the high temperature regime $T = 0.9 T_c$ (red line in Fig.~\ref{fig:figs5}b). As such a dramatic enhancement in $B_{c2}$ was not observed in the experiment, we believe the intra-orbital $A_g$ pairing we assumed in the main text provides a more plausible description of the superconducting state in monolayer WTe$_2$.  

We further note that with the $A_g$ intra-orbital singlet pairing discussed in the main text, the effective BdG model under the pseudospin basis describes an $s$-wave superconductor with even-parity pairing, which is known to be topologically trivial and thus distinct from the odd-parity pairing studied in Cu-doped Bi$_2$Se$_3$ \cite{FuliangS}. In this trivial $A_g$ phase, the quantum spin Hall edge states will acquire a full pairing gap and no helical Majorana edge modes can form. However, by placing a ferromagnetic insulator to cover half of the superconducting quantum spin Hall insulator, a Majorana fermion can form at the ferromagnet-superconductor interface \cite{FuKane}.

As we explained in the subsections above, the odd-parity $B_u$ pairing has a similar topological nature as the odd-parity pairing studied in Cu-doped Bi$_2$Se$_3$. However, the edge states will still be gapped out when the disconnected $Q$-valleys are filled and no TRIM point is enclosed by the Fermi surface, as shown explicitly in Fig.\ref{fig:figs6}c. When the Fermi surface encloses an odd number of TRIM points, the chemical potential is generally lying deep in the bulk bands. In this case, the quantum spin Hall edge states have already merged deeply into the bulk and do not participate in the edge physics. However, since the superconducting phase is topological, helical Majorana mode will emerge in this DIII class topological superconductor as shown in Fig.\ref{fig:figs6}f.

\subsection{Four-band tight-binding model for superconducting WTe$_2$}

In this subsection, we present details of the tight-binding model used to study the bulk-edge correspondence in the nontrivial $B_u$ phase in Fig.\ref{fig:figs6}. In the Nambu basis $(c_{\bm{k},p,\uparrow},c_{\bm{k},p,\downarrow}, c_{\bm{k},d,\uparrow},c_{\bm{k},d,\downarrow}, c^{\dagger}_{-\bm{k},p,\uparrow}, c^{\dagger}_{-\bm{k},p,\downarrow}, c^{\dagger}_{-\bm{k},d,\uparrow}, c^{\dagger}_{-\bm{k},d,\downarrow})^{T}$, where $c^{\dagger}_{\bm{k},l,\sigma}$ ($l=p,d,\sigma = \uparrow, \downarrow$) creates a Bloch state formed by linear combinations of Wannier orbital of character $l$ and spin $\sigma$, the momentum-space tight-binding Hamiltonian $\hat{H}^{TB}_{BdG}(\bm{k})$ for superconducting monolayer WTe$_2$ under $\hat{\Delta}_1$ reads:
\begin{eqnarray}
\hat{H}^{TB}_{BdG}(\bm{k}) &=& \sum_{\bm{k}, mn} c^{\dagger}_{\bm{k},m} H^{TB}_{0, mn} (\bm{k}) c_{\bm{k},n} + \Delta_1 (c^{\dagger}_{\bm{k},p,\uparrow} c^{\dagger}_{-\bm{k},d,\downarrow} - c^{\dagger}_{\bm{k},p,\downarrow} c^{\dagger}_{-\bm{k},d,\uparrow} + h.c.).
\end{eqnarray}
Here, $m,n = (l, \sigma)$ label the index for different Wannier orbitals with $l= p,d$, $\sigma=\uparrow, \downarrow$. $H^{TB}_{0} (\bm{k})$ is a $4\times4$ matrix given by:
\begin{eqnarray}\label{eq:HTB}
H^{TB}_{0} (\bm{k})=
\begin{pmatrix}
E_p (\bm{k}) - \mu & 0 & -i v_0\sin(k_x a) + \alpha_z\sin(k_y b) & -i\alpha_x\sin(k_x a) + \alpha_y\sin(k_y b)\\
  & E_p (\bm{k}) - \mu & i\alpha_x\sin(k_x a) + \alpha_y\sin(k_y b) & -i v_0\sin(k_x a) - \alpha_z\sin(k_y b)\\
 &   & E_d(\bm{k})  - \mu   & 0\\
 h.c.&  &  &     E_d(\bm{k}) - \mu
\end{pmatrix},
%\begin{pmatrix}
%E_p (\bm{k}) - \mu & 0 & -i v_0\sin(k_x a) + \alpha_z\sin(k_y b) & -i\alpha_x\sin(k_x a) + \alpha_y\sin(k_y b)\\
% 0 & E_p (\bm{k}) - \mu & i\alpha_x\sin(k_x a) + \alpha_y\sin(k_y b) & -i v_0\sin(k_x a) - \alpha_z\sin(k_y b)\\
%i v_0\sin(k_x a) + \alpha_z\sin(k_y b) & -i\alpha_x\sin(k_x a) + \alpha_y\sin(k_y b) & E_d(\bm{k})  - \mu   & 0\\
%i\alpha_x\sin(k_x a) + \alpha_y\sin(k_y b) & i v_0\sin(k_x a) - \alpha_z\sin(k_y b) & 0 &                E_d(\bm{k}) - \mu
%\end{pmatrix}
\end{eqnarray}
where 
\begin{eqnarray}
E_p (\bm{k}) &=& 2 t_{1p} \cos(k_x a) + 2 t_{2p}\cos(k_y b) - u_p - 2(t_{1p} + t_{2p}), \\\nonumber
E_d(\bm{k}) &=& 2 t_{1d}\cos(k_x a) + 2 t_{2d}\cos(k_y b) + 2 t'_{2d}\cos(2 k_y b) - u_d - 2 (t_{1d} + t_{2d} + t'_{2d}).
\end{eqnarray}
The parameters in $H^{TB}_{0} (\bm{k})$ above are tabulated in Table \ref{tableTB} below. It can be verified in a straightforward way that $H^{TB}_{0} (\bm{k})$ reduces to the $\bm{k} \cdot \bm{p}$ model near the $\Gamma$-point in Eq.\ref{eq:H2} in the continuum limit $a,b \rightarrow 0$.

To obtain the edge spectrum in Fig.\ref{fig:figs6}, we perform partial Fourier transform: $c_{k_y, m}(x) = (1/\sqrt{L_x}) \sum_{k_x} e^{-i k_x x} c_{\bm{k}, m}$ and set open boundary conditions for edges terminated at $x = 0$ and $x = L_x = 400$. In Fig.\ref{fig:figs6}a-c, we set $\mu = 60$ meV such that two Q-pockets form at the Fermi surface. In Fig.\ref{fig:figs6}d-f, we set $\mu = -0.6$ eV such that the hole band is accessed with a single $\Gamma$-pocket, and the paring amplitude for $\hat{\Delta}_1$ in Fig.\ref{fig:figs6}f is set to be $\Delta_1=|\mu|/10$.

\begin{table}[h]
	\centering
	\caption{Tight-binding parameters in $H^{TB}_{0} (\bm{k})$(Eq.\ref{eq:HTB}) in units of eV. Lattice constants: $a=6.31${\AA}, $b=3.49${\AA}.}
	\begin{tabular}{ccccccccccc}
		\hline\hline
		$u_p$ & $u_d$ & $t_{1p}$ & $t_{2p}$ & $t_{1d}$ & $t_{2d}$ &$t'_{2d}$ & $v_0$ &$\alpha_x$ & $\alpha_y$ & $\alpha_z$ \\
		\hline
		-1.39 & 0.062 & 0.626 & 1.517 & -0.06 & -0.387 & 0.15 & 0.371 & 0.027 &0.163 &0.020 \\
		\hline
	\end{tabular}	
	\label{tableTB}
\end{table}

\end{document}